\documentclass[12pt,a4paper]{article}
\usepackage{amsmath}
\usepackage{amsthm}
\usepackage{amsfonts}
\usepackage{amssymb}
\usepackage{graphicx}
\usepackage{color}
\definecolor{darkgreen}{rgb}{0,0.5,0}
\definecolor{purple}{rgb}{1,0,1}
\newcount\Comments  %
\newcommand{\kibitz}[2]{\ifnum\Comments=1\textcolor{#1}{#2}\fi}

\numberwithin{equation}{section}
\usepackage{srcltx}
\usepackage{authblk}
\usepackage[left=2.00cm, right=2.00cm, top=2.00cm, bottom=2.00cm]{geometry}
\theoremstyle{definition}
\newtheorem{exmp}{Example}[section]
\begin{document}
\title{On Rational Solutions of Dressing Chains of Even Periodicity}

\author[1]{ H. Aratyn}
\author[2]{J.F. Gomes} 
\author[2]{G.V. Lobo} 
\author[2]{A.H. Zimerman}
\affil[1]{
Department of Physics, 
University of Illinois at Chicago, 
845 W. Taylor St.
Chicago, Illinois 60607-7059, USA}
\affil[2]{
Instituto de F\'{\i}sica Te\'{o}rica-UNESP,
Rua Dr Bento Teobaldo Ferraz 271, Bloco II,
01140-070 S\~{a}o Paulo, Brazil}

\maketitle

\abstract{
We develop a systematic approach to deriving  rational
solutions and obtaining classification  of their parameters
for dressing chains of even $N$ periodicity or equivalently
 Painlev\'e  equations invariant under $A^{(1)}_{N-1}$ symmetry.  
This formalism identifies rational solutions (as well as special  function solutions)
with points on orbits of fundamental shift operators of $A^{(1)}_{N-1}$ affine Weyl group  acting on seed configurations 
defined as  first-order polynomial solutions of the underlying dressing chains.
This approach clarifies the structure of rational solutions and establishes explicit and systematic method towards their construction.

For the special case of the $N=4$ dressing chain equations  the 
method yields all the known rational (and special function) solutions of Painlev\'e V equation.
The formalism naturally extends to $N=6$ and beyond as shown in
the paper.
}

\section{Introduction and background information}
\label{section:intro}

Painlev\'e equations form a class of second order nonlinear differential equations
with solutions  that  have no movable critical singularities in the complex plane, see e.g.  \cite{gromak}. 
Although this mathematical property motivated the discovery of Painlev\'e equations, in relatively short time,  these equations made astonishing impact on several  fields inside and outside mathematics.
A long and incomplete list of affected topics and models includes 
correlation functions of the Ising model, random matrix theory, plasma physics, nonlinear waves, quantum gravity, quantum ﬁeld theory, general relativity, nonlinear and ﬁber optics, and  Bose-Einstein condensation.
Special solutions, such as rational solutions, turned out to be 
important in these applications and various methods were applied in their study.
To provide a systematic approach to the study of rational solutions we here utilize  
the dressing chain and its  connection to Painlev\'e equations. The dressing chain
was derived by applying Darboux transformations to the 
spectral  problem of second order differential equation \cite{veselov}. 
Specifically, let us  consider a sequence of second order 
differential operators $L_n$  connected
via first order Darboux transformations :
$ (\partial_z- j_n) L_n = (L_{n-1} +\alpha_n) (\partial_z- j_n)$, 
where $\alpha_n$ is a constant. 
Such symmetry is realized for 
\begin{equation} 
L_n = (\partial_z+ j_n) (\partial_z- j_n) +\alpha_n =
 (\partial_z- j_{n+1}) (\partial_z+ j_{n+1})\, , 
\label{darboux}
\end{equation}
with $L_n$ defined by products of two first order differential operators with their
orders being interchanged when going from $n$ to $n+1$.
Comparing the two alternative expressions for $L_n$ in equation \eqref{darboux}
we obtain the nonlinear lattice equations  \cite{veselov} :
\begin{equation}
(j_n+j_{n+1} )_z = -j_n^2+j_{n+1}^2 +\alpha_n, \; n=1,{\ldots} , N,
\quad j_{N+i}=j_i\, ,
\label{dressingeqs}
\end{equation}
made finite by imposing the periodic boundary condition $j_{N+i}=j_i$.
We refer to system \eqref{dressingeqs} as a system of dressing 
chain equations of $N$-periodicity.
Such system possesses many important properties. For $N = 3$, it has
been shown \cite{veselov} that it passes the Kovalevskaya-Painlev\'e 
test and its equivalence to Painlev\'e IV equation has also been established
\cite{adler,veselov}.  For higher $N$ the system is equivalent to 
$A^{(1)}_{N-1}$ invariant  Painlev\'e  equations \cite{adler,noumi98} and this equivalence 
will be utilized  in this paper to construct and study rational solutions of Painlev\'e equations in the context of underlying periodic dressing chains.
 Quite recently the $N$ cyclic dressing chain was also obtained in  
the self-similarity limit of the second flow of $sl(N )$ mKdV hierarchy
\cite{victor}.

As we will now show the system \eqref{dressingeqs} requires different treatments
depending on whether $N$ is odd or even.
This becomes evident when we consider a regular sum 
$ \sum_{n=1}^N (j_n+j_{n+1} )_z $ and an alternating sum 
$ \sum_{n=1}^N (-1)^n (j_n+j_{n+1} )_z $ of derivatives of $j_n+j_{n+1}$.
Calculating a regular sum using
the dressing equations \eqref{dressingeqs} we obtain for both 
even and odd $N$ the same expression
\begin{equation}
 \sum_{n=1}^N (j_n+j_{n+1} )_z = 2  \sum_{n=1}^N (j_n)_z = \sum_{n=1}^N
 \alpha_n \, ,
\label{vinculo}
\end{equation}
 for the integration constant on the right hand side.
As long as $N$ is odd calculating an alternating sum  $ \sum_{n=1}^N (-1)^n
(j_n+j_{n+1} )_z $ using the dressing equations \eqref{dressingeqs}
will reproduce the same  condition as in  \eqref{vinculo}.
For  even $N$ the alternating sum $ \sum_{n=1}^N (-1)^n
(j_n+j_{n+1} )_z $ is identically zero (positive and negative terms
simply cancel). However the same expression
calculated by plugging the right hand side of 
dressing equations \eqref{dressingeqs} yields
for e.g. $N=4$ the expression  $\left( j_1^2+j_3^2-j_2^2-j_4^2 \right)  
+\frac12 (-\alpha_1 +\alpha_2-\alpha_3+\alpha_4 )\,$. 
Thus the dressing chains of even periodicity require imposition of a
new quadratic constraint or modification of the dressing chain
formulation. Such modification was proposed in 
\cite{AGZ2021}, where the authors put forward  
a system of dressing chain  equations 
of even $N =4, 6 , 8 ,{\ldots}  $ periodicity  defined as :
\begin{equation}
(j_i+j_{i+1} )_z = -j_i^2+j_{i+1}^2 +\alpha_i+(-1)^{i+1}
\frac{(j_i+j_{i+1}) \Psi}{\Phi}\
, \; i=1,2, {\ldots}
, N, \quad j_{N+i}=j_i\, ,
\label{dressingeqseven}
\end{equation}
where 
\begin{equation}
 \Psi=  \sum_{k=1}^N (-1)^{k+1} \left( j_k^2
- \frac12  \alpha_k \right), \qquad \Phi=\sum_{k=1}^N j_k \, .
\label{psidef}
\end{equation}
This structure is such that both regular and  alternating sums
of derivatives of $j_i+j_{i+1}$ give 
consistent answers when applied on 
the system \eqref{dressingeqseven}:
\[ \begin{split} \sum_{i=1}^N (j_i+j_{i+1} )_z& = 2 \Phi_z = 
\sum_{i=1}^N \alpha_i\, , 
\\
\sum_{i=1}^N (-1)^i (j_i+j_{i+1} )_z& =2
\sum_{k=1}^N (-1)^{k+1}  j_k^2   
+\sum_{k=1}^N (-1)^{k}\alpha_k - 2 \frac{\Phi}{\Phi}
\Psi= 0 \, .
\end{split}\]

As shown in \cite{AGZ2021} such system can be obtained  by Dirac
reduction from $N+1$ dressing chain \eqref{dressingeqs} of odd
periodicity.  

The above equations as well as quantities $\Psi$ and $\Phi$
are invariant under $A^{(1)}_{N-1}$  B\"acklund transformations
$ s_i,\, i=1,{\ldots} , N$ \cite{adler} :
\begin{equation}
 j_i \stackrel{s_i}{\longrightarrow}
  j_i - \frac{\alpha_i}{j_i
+j_{i+1}},\;\;
j_{i+1}   \stackrel{s_i}{\longrightarrow}  
 j_{i+1} + \frac{\alpha_i}{j_i
+j_{i+1}}, \;\;
j_k \stackrel{s_i}{\longrightarrow} j_k, \; k \ne i, k\ne i+1\, ,
\label{Tinewj}
\end{equation}
when transformations \eqref{Tinewj} are accompanied by transformations
of coefficients 
\begin{equation} \alpha_i \to
-\alpha_i, \quad \alpha_{i \pm 1}\to \alpha_{i \pm 1} +\alpha_i.
\label{sialphaj}
\end{equation}

There are also two automorphisms $\pi, \rho$ :
\begin{equation} \begin{split}
\pi  &:  j_i \to j_{i-1}, \; \alpha_i \to \alpha_{i-1}, \;
\pi(\Phi)=\Phi, \; \pi(\Psi)=-\Psi \\
\rho &:  z\to -z , j_i \rightarrow -j_{i+2},\;\;
\alpha_i \rightarrow \alpha_{i+2},\;
\; \rho(\Phi)=-\Phi, \; \rho(\Psi)=\Psi \, ,
\label{dressauto}
\end{split}
\end{equation}
that keep the dressing equations \eqref{dressingeqseven} invariant.

For the redefined quantities
\begin{equation}
{\bar j}_{n}=j_n +\frac{(-1)^{n}}{2} \frac{\Psi}{\Phi}\, , 
\label{jbardef}
\end{equation}
it holds that the corresponding sum  
$f_n = j_n+j_{n+1}={\bar j}_{n}+{\bar j}_{n+1}$ is unchanged.
Such redefinition leads to a formal absorption of $\Psi$ terms so that
they are no longer explicit in the dressing equations 
rewritten in terms of ${\bar j}_{n}$ that satisfy equations 
\eqref{dressingeqs} \cite{AGZ2021}. However such process introduces
potential extra divergencies 
into an associated Sturm-Liouville problem.
Throughout this paper we will work with 
\eqref{dressingeqseven} with a constant non-zero $\Psi$ so that 
the polynomial seed solutions we will construct below will be free of divergencies.

We present construction of rational and special function
solutions for  dressing chains of even periodicity. 
In this work rational solutions are identified with points on the
orbits of fundamental shift operators (sometimes also referred to in the literature as translations) of the extended affine Weyl group $A^{(1)}_{N-1}$ acting on the first-order polynomial
seed solutions. 
In particular for the seed solutions with all the 
components being equal to each other the
construction yields rational solutions being ratios of Umemura
polynomials \cite{noumi98u}.
The reduction procedure that yields special function solutions is outlined 
and is shown to reproduce rational solutions for appropriate values of
the parameters of the underlying Riccati equations. 

The presentation is organized as follows. 
In section \ref{section:polynomials}, we obtain the first-order polynomial
solutions of the dressing chain equations \eqref{dressingeqseven}
with parameters $\alpha_i, i=1,{\ldots} ,  N$ 
depending on one arbitrary variable and with a constant non-zero
$\Psi$ that ensures that the solution is polynomial.

In section \ref{section:hamiltonian}, we establish connection between
 the dressing chain equations \eqref{dressingeqseven} and
 Hamiltonian formalism for $N=4,6$ that can easily be generalized to
 arbitrary even $N$.
Essential for establishing this connection is 
ability to cast the dressing chain equations \eqref{dressingeqseven} as
symmetric $A^{(1)}_{N-1}$-invariant Painlev\'e equations as those
given in equations \eqref{N4feqs} and  \eqref{N6NY} for $N=4,6$,
respectively. 
We should point out that translating the system of equations
depending on $j_i$ into formalism that is expressed entirely in terms
of $f_i=j_i+j_{j+1}$  is possible for even  $N$  thanks due to the
presence of $\Psi$ terms on the right hand sides  of 
equations \eqref{dressingeqseven}.
This is in contrast to odd $N$
dressing chains  where 
$j_i$ and $f_i$ are always fully interchangeable.
For $N=4$ the Hamiltonian formalism of section \ref{section:hamiltonian}
gives rise to Painlev\'e V equation as  briefly  reviewed in subsection
\ref{subsection:H2pain}. The first-order polynomial
solutions in the setting of Hamiltonian formalism become the algebraic solutions
of \cite{watanabe}. 

We are able to present a power series expansions 
of Hamiltonian variables $p$ and $q$ in subsection \ref{subsection:kova}.
We show  how potential  divergencies of power series 
solutions (that can not be absorbed in $\Psi$) can be removed by
appropriate B\"acklund transformations.  
After removing the eventual simple poles from rational solutions by 
acting with the B\"acklund transformations we obtain rational solutions 
that are expandable in a series of positive powers of $z$ and can be
reproduced by actions of the shift operators as shown in the next
section.

In section \ref{section:rational}, we derive rational solutions for
$N=4$ by acting 
with shift operators on the first-polynomial solutions \eqref{solution2}
and \eqref{solution1} to obtain all known cases listed in reference
\cite{kitaev}
that presented necessary and sufficient conditions for rational solutions of 
Painlev\'e V equation. 
Reference \cite{ohta} showed how to act with shift operators on
solutions \eqref{solution2} (expressed by tau functions) to obtain 
some of the cases of \cite{kitaev} (items I + II on page 
\pageref{page:items}). 
For the first-order polynomial seed solutions  \eqref{solution2} 
(with all the components $j_i$ equal to $z/N$) the action of shift operators 
yields rational solutions expressed by Umemura polynomials
\cite{noumi98u,kajiwara} and we use the shift
operators to derive the recurrence relations that determine these polynomials.
Extending structure of seed solutions to
include  solutions \eqref{solution1} (where $j_i+j_{i+1}=0$ for some
$i$)
requires exclusion of those shift operators
that are ill-defined when acting on such solutions as discussed in
subsection \ref{subsection:item3}. Those of the 
shift operators that are well-defined generate the remaining 
rational solutions  from solutions \eqref{solution1}, see item III on page 
\pageref{page:items1}. This new approach leads to a systematic 
and unified way to derive all rational Painlev\'e V solutions.
Based on results for $N=4$ we conjecture for all even $N$ that all rational
solutions are obtainable through actions of shift
operators on first-order polynomial solutions.

In section \ref{section:N=6},  we provide explicit
construction of special function solutions   and rational solutions for $N=6$.
The rational solutions are always identified with orbits of the
fundamental shift operators. 
For the seed solution  with all components being equal 
or 
only one of the components being negative we are able to express 
the corresponding rational solutions by Umemura type of polynomials.
Existence of special function solutions
is established for the remaining cases  with a sufficient number of constraints 
imposed on $\alpha_i$ parameters to insure  reduction of Hamiltonian equations 
to one single Riccati equation. For $N=6$ case this happens for three 
independent constraints. However we also encounter hybrid situations with one single Riccati equation and
one coupled quadratic (in $q_i,p_i$) equation for some cases with
two constraints. In such cases there exists a special function solution 
for only one of the variables. Interestingly, when $\alpha_i$ parameters are
associated with orbits of the shift operators we obtain closed expressions 
in terms of Whittaker functions that describe rational solutions for 
all underlying variables of the reduced system.

\section{Preliminaries. The seed solutions as the first-order polynomial solutions of even chains}
\label{section:polynomials}
For simplicity we first carry out the discussion for $N=4$ before
proceeding
to the case of $N=6$ and making  general comments about higher $N$ cases.

We are looking for the first-order polynomial solutions to equation
\eqref{dressingeqseven}  of the type
\[
j_i= c_i z ,\quad \sum_{i=1}^4 c_i =1 \, , 
\]
that satisfy  the $\Phi=z$ condition. With such ansatz the quantity 
$\Psi$ defined in \eqref{psidef} can
only contain terms with $z^2$ or a constant. The terms  quadratic in $z$
can be absorbed in $j_i$ via \eqref{jbardef} transformation. Thus without
losing any generality we can assume that 
\begin{equation} \Psi= \frac12 (-\alpha_1 +\alpha_2-\alpha_3+\alpha_4
)= 1-\alpha_1 -\alpha_3\,,
\label{constpsi}
\end{equation}
where we used that $\sum_{i=1}^4 \alpha_i=2$.

One can easily see that the condition for $\Psi$  not to  contain $z^2$
for the polynomial solutions of the first-order amounts to
$j_{n+1}^2-j_n^2=0$ on the right hand side of the dressing equations. Thus
the solution must be $j_i=z c (\epsilon_1,
\epsilon_2,\epsilon_3,\epsilon_4)$
with $\epsilon_i=\pm 1$ and $c$ a non-zero constant. Since $\Phi=z\ne0$ we 
must also have $\epsilon_1+\epsilon_2+\epsilon_3+\epsilon_4 \ne0$.
This argument eliminates the case of two epsilons being negative,
$\epsilon_i=-1, \epsilon_j=-1, i\ne j$,  as this would violate $\Phi \ne 0$.
Therefore the  only two independent (up to $\pi$) polynomial solutions
are :
\begin{align}
&j_i= \frac{z}{4} (1,1,1,1)& 
&( \alpha_1, \alpha_2, \alpha_3, \alpha_4)= 
(\mathsf{a},1-\mathsf{a},\mathsf{a}, 1-\mathsf{a}) &  
& \Phi=  z , \Psi=1-2 \mathsf{a}  \, , 
\label{solution2} \\
& j_i= \frac{z}{2} (1,1,-1,1) & 
&  (\alpha_1, \alpha_2, \alpha_3, \alpha_4) =
(\mathsf{a},0,0,2-\mathsf{a}) &
&\Phi=z, \Psi=1-\mathsf{a}  \, .
\label{solution1} 
\end{align}
Both solutions depend on only one free parameter $\mathsf{a}$. 
The remaining first order polynomial solutions can be obtained by
acting with  $\pi, \pi^2$ and $\pi^3$  on solution \eqref{solution1} (recall that $\pi^4=1$ for $N=4$
cyclicity and so $\pi^3=\pi^{-1}$). Note that in case of solution
\eqref{solution1} the action of automorphism $\pi$ is such that it
simply moves the 
$-1$ term in expression for $j_i$ and zeros in expression for $\alpha_i$ 
to the right. It is important to point out that there could be other potential solutions of the first-order polynomial type like
for example $j_i=(z/2) (1,0,1,0)$. However such solutions would involve $z^2$ terms in $\Psi$ and could be transformed  by the 
transformation \eqref{jbardef} involving the $z^2$ part of $\Psi$ to the solution \eqref{solution1} or its $\pi$ variants.

One can easily extend this analysis to higher $N$  with $\Psi$ and
$\Phi$ defined in the definition \eqref{psidef}.
For the $N=6$   first-order polynomial solutions we take :
\[
\Psi= \frac12 (-\alpha_1 +\alpha_2-\alpha_3+\alpha_4 -\alpha_5+\alpha_6)=
1-\alpha_1 -\alpha_3 -\alpha_5\, ,
\]
and obtain five different first-order polynomial solutions:
\begin{align}
j_i&=\frac{z}{6} (1,1,1,1,1,1), \; \quad
\alpha_i=(\mathsf{a},\frac23-\mathsf{a},\mathsf{a}, \frac23-\mathsf{a}
,\mathsf{a}, \frac23-\mathsf{a}) \, , \label{N6polynomialsols1}\\
j_i&=\frac{z}{4} (1,1,1,1,1,-1), \;\quad
\alpha_i=(\mathsf{a},1-\mathsf{a},\mathsf{a}, 1-\mathsf{a} , 0,0)
\, , \label{N6polynomialsols2}\\
j_i&=\frac{z}{2} (1,1,1,1,-1,-1), \;\quad
\alpha_i=(\mathsf{a},2-\mathsf{a}, \mathsf{a}
,0,-\mathsf{a},0)\, ,\label{N6polynomialsols3}\\
j_i&=\frac{z}{2} (1,1,1,-1,1,-1),\; \quad
\alpha_i=(2-\mathsf{a},\mathsf{a}, 0
,0,0,0)\, ,\label{N6polynomialsols4}\\
j_i&=\frac{z}{2} (1,1,-1,1,1,-1)
,\; \quad \alpha_i=(2-\mathsf{a}, 0,0,\mathsf{a},0,0)
\, ,\label{N6polynomialsols5}
\end{align}
since all these configurations seems to be distinct and can not be
connected by permutation generated by $\pi$ or multiples
of $\pi$'s. All the above solutions depend on one arbitrary parameter
$\mathsf{a}$.
Note that $j_i=z (1,1,1,-1,-1,-1)$ is not a solution because it
would violate $\Phi \ne0$ condition. Thus the number of configurations
is equal to $1+1 +p(6-2, 2)=5$, where $p(6-2, 2)=p(4, 2)=3$ is a number of
partitions of $4$ in two parts (of positive integers and zero) : $4=4+0=3+1=2+2$.
For $N=8$ we find a number of the first-order polynomial
solutions to be $1+1+ p(8-2,2)+p(8-3,3)$ with 
$p(8-2,2)=p(6,2)=4$ and $p(8-3,3)=p(5,3)=5$. Generally 
a number of the first-order polynomial
solutions is given by $1+1+\sum_{k=2}^{N/2-1} p(N-k,k)$ where
$p(N-k,k)$ is a number of distinct partitions of $N-k$ in $k$ parts consisting of positive
integers and zero. 

For arbitrary even $N$ with $\Phi =z$, $\Psi= 1 - \sum_{k=1}^{N/2}
\alpha_{2k-1}$  and an arbitrary variable $\mathsf{a}$ 
there will always be a fully symmetric solution:
\[ j_i= \frac{z}{N}, \; i=1,{\ldots} ,N, \quad  \alpha_{2j-1}=
\mathsf{a},\; \alpha_{2j}=\frac{4}{N}-\mathsf{a},\;  j=1,{\ldots}
,N/2\, ,
\]
which is a fixed point of $\pi^2$ automorphism.
The remaining solutions will have one and up to $N/2-1$ 
negative components $j_i= -\frac{z}{N}$ with varying distance between
the negative components. For example for only one negative component
in the last position we get
\[j_i= \frac{z}{N-2}, \; i=1,{\ldots}, N-2,\;
\alpha_{2j-1}=
\mathsf{a},\; \alpha_{2j}=\frac{4}{N-2}-\mathsf{a},\;  j=1,{\ldots}
,N/2-1\, ,
\]
with  $j_k=0, \alpha_k=0$ for $k=N-1,N$, and so on for solutions with
more negative components.

One needs to point out that the first-order solutions 
\eqref{N6polynomialsols1}-\eqref{N6polynomialsols5} 
appeared also as simple rational solutions expressed in terms of
$f_i=j_i+j_{i+1}$ that give rise to other rational solutions via
B\"acklund transformations in the framework of $A^{(1)}_5$ 
Painlev\'e equations (equivalent to $N=6$ dressing chain equations)
in reference \cite{matsuda5}. 

\section{Hamiltonian formalism and polynomial solutions}
\label{section:hamiltonian}

\subsection{Hamilton equations and their algebraic solutions}
For $N=4$ we will show how the first-order polynomial solutions 
\eqref{solution2} and \eqref{solution1} are equivalent to all
algebraic solutions found for Painlev\'e V equation in \cite{watanabe}. 
These solutions will then  serve as seeds of all rational solutions \cite{kitaev}
of  Painlev\'e V equation via shift transformations.

Thanks to the presence of $\Psi$ in the dressing equations 
\eqref{dressingeqseven} they can be rewritten  in terms $f_i=j_i+j_{i+1}, i=1,2,3,4$ 
as 
\begin{equation}
\begin{split}
z \frac{d f_1}{d z}&=
f_1 f_3 \left(f_2-f_4\right) +
\left(1-\alpha_3\right) f_1 + \alpha_1 f_3 \, , \;
z \frac{d f_2}{d z}=
f_2 f_4 \left(f_3-f_1\right) +\left(1-\alpha_4\right) f_2 + \alpha_2 f_4
 ,\\
 z \frac{d f_3}{d z}&= f_1 f_3 \left(f_4-f_2\right) +
\left(1-\alpha_1\right) f_3 + \alpha_3 f_1 \, ,\;
z \frac{d f_4}{d z}= f_4 f_2 \left(f_1-f_3\right) +
\left(1-\alpha_2 \right) f_4 + \alpha_4 f_2 \, ,
\label{N4feqs}
\end{split}
\end{equation}
after multiplication by $\Phi=f_1+f_3=f_2+f_4=z$ 
and use of definition of $\Psi$ from \eqref{psidef}. Recall 
that it follows from relation \eqref{vinculo} that
$\alpha_1+\alpha_2 +\alpha_3+\alpha_4=2$.

The above system of equations can be cast into a Hamiltonian system with
\begin{equation}
H= -q\,(q-z)\,p\, (p-z)+(1-\alpha_1-\alpha_3)\, p q +\alpha_1 z
p-\alpha_2 z q \, ,
\label{pqHam}
\end{equation}
with Hamilton equations 
\begin{equation}\begin{split}
z q_z &= -q(q-z)(2p-z)+(1-\alpha_1-\alpha_3)  q +\alpha_1 z \, ,\\
z p_z &= p (p-z)(2q-z) - (1-\alpha_1-\alpha_3) p +\alpha_2 z \, ,
\label{hameqs}
\end{split}
\end{equation}
derived from 
\[ z q_z = \frac{d H}{d p} , \qquad  z p_z =- \frac{d H}{d q}\,.\]

The Hamilton equations \eqref{hameqs} reproduce the $N=4$ system
of equations \eqref{N4feqs} after substitution $(q,p) \to (f_1,f_2,f_3,f_4)$ such that
\[
q= f_1=z-f_3, \qquad p =f_2=z-f_4\, .
\]
The B\"acklund transformations \eqref{Tinewj} and automorphisms
\eqref{dressauto} 
are given in the  setting of Hamilton equations \eqref{hameqs} by
\begin{equation}
\begin{split}
s_1&: q \to q, \; p \to p+ \frac{\alpha_1}{q}, \; \alpha_1 \to -\alpha_1, \,
\alpha_2 \to \alpha_2+\alpha_1, \alpha_3 \to \alpha_3\\
s_2&: q \to q- \frac{\alpha_2}{p}, \; p \to p, \; \alpha_2 \to -\alpha_2, \,
\alpha_1 \to \alpha_2+\alpha_1, \alpha_3 \to \alpha_3+\alpha_2\\
s_3&: q \to q, \; p \to p -\frac{\alpha_3}{z-q}, \; \alpha_3 \to
-\alpha_3, \,
\alpha_2 \to \alpha_2+\alpha_3, \alpha_4 \to \alpha_3+\alpha_4\\
s_4&: q \to q+ \frac{\alpha_4}{z-p}, \; p \to p, \; \alpha_4 \to
-\alpha_4, \,
\alpha_1 \to \alpha_4+\alpha_1, \alpha_3 \to \alpha_3+\alpha_4\\
\pi  &:  q \to z-p, \; p \to q ,\;  
\alpha_i \to \alpha_{i-1},  \\
\rho &:  z\to -z ,\; q \to z-q ,\; p \to p-z,  \;
\alpha_1 \leftrightarrow \alpha_3,\; \alpha_2 \leftrightarrow \alpha_4,
\label{backsymham}
\end{split}
\end{equation}
where $\alpha_4$ is understood as $2-\alpha_1-\alpha_2-\alpha_3$
in terms of $\alpha_i, i=1,2,3$ appearing in the Hamiltonian formalism.

Solutions \eqref{solution2} and \eqref{solution1} as well solutions
that can be obtained from \eqref{solution1} by an automorphism $\pi : j_i \to
j_{i-1}, \alpha_i \to \alpha_{i-1}$  are given in terms of $q,p$ by
\begin{align}
q&=z/2, \; p=z/2, \; (\mathsf{a}, 1-\mathsf{a}, \mathsf{a}, 1-\mathsf{a})\, ,
\label{wata3} \\
q&=z, \;p=0, \; (\mathsf{a}, 0,0, 2-\mathsf{a}) \, ,  
\label{wata1}\\
q&=z, \; p=z, \;(\mathsf{a}, 2-\mathsf{a},0,0)\, , 
\label{wata2}\\
q&=0, \;  p=z,\;  (0,\mathsf{a}, 2-\mathsf{a},0)  \label{wata5} \, ,\\
q&=0,\; p=0, \; (0,0, \mathsf{a},2-\mathsf{a})  \label{wata6} \, ,
\end{align}
where \eqref{wata3} is derived from  \eqref{solution2} while the
remaining solutions are obtained from  \eqref{solution1} and its $\pi$
variants. Solution \eqref{wata3} is a fixed point of $\pi^2$ while all the remaining solutions can be connected to each other 
by the $\pi$ automorphism.
All these solutions coincide with a  
complete set of algebraic solutions found by Watanabe \cite{watanabe}.

For $N=6$ we define the Hamiltonian formalism in terms of
quantities:
\begin{equation}
q_1=j_1+j_2,\;\;\; p_1= j_2+j_3,\;\;\;q_2=j_1+j_2+j_3+j_4,\;\;\; p_2=j_4+j_5\, ,
\label{qipij}
\end{equation}
which satisfy equations
\begin{equation}
\begin{split}
z q_{1, \, z}&= q_1(q_2-q_1)(2 p_1-z)+q_1 (z-q_2)(2p_1+2p_2-z) 
+z \alpha_1 +q_1 (1-\alpha_1-\alpha_3-\alpha_5)\\
z q_{2, \, z}&= (q_2-q_1)(z-q_2)(2p_2-z)+
q_1(z-q_2)(2p_1+2p_2-z)
+z (\alpha_1+\alpha_3) +q_2 (1-\alpha_1-\alpha_3-\alpha_5)\, ,\\
z p_{1, \, z}&= p_1 p_2 (2 q_2 -2 q_1-z)+p_1 (z-p_1-p_2)(z-2q_1) 
+z \alpha_2 -p_1 (1-\alpha_1-\alpha_3-\alpha_5)\\
z p_{2, \, z}&= p_1 p_2 (2 q_1-2 q_2+z)+p_2  (z-p_1 -p_2)(z-2 q_2) 
+z \alpha_4 -p_2 (1-\alpha_1-\alpha_3-\alpha_5)\, ,
\end{split}
\label{q1p2eqs}
\end{equation}
that can be derived from  $N=6$ dressing chain \eqref{dressingeqseven}
(explicitly given for $N=6$ in the appendix in equation \eqref{N6NY}).
Equations \eqref{q1p2eqs} can be realized as Hamilton equations
$z q_{i\, z}= \partial H /\partial p_i$ and 
$z p_{i\, z}= -\partial H /\partial q_i$ for $i=1,2$ with the
Hamiltonian :
\[ \begin{split} H&= -\sum_{i=1}^2 p_i(p_i-z)q_i (q_i-z) -2 p_1 q_1 p_2 (q_2-z)
+\sum_{i=1}^2 p_i z \sum_{j=1}^i \alpha_{2i-1} 
-\sum_{i=1}^2 q_i z \alpha_{2i}\\& + 
\sum_{i=1}^2 q_i p_i (1-\alpha_1-\alpha_3-\alpha_5)\, .
\end{split}\]
One of advantages of variables $q_i,p_i, i=1,2$ is that they make
expressions for B\"acklund transformations \eqref{Tinewj}  more
transparent. The actions of B\"acklund transformations on these
variables are given by
\begin{equation}
\begin{split}
s_1(p_1) &= p_1+\frac{\alpha_1}{q_1},\; s_2(q_1)=q_1-\frac{\alpha_2}{p_1}
,\; s_3(p_1)=p_1-\frac{\alpha_3}{q_2-q_1} ,\;
s_3(p_2)=p_2+\frac{\alpha_3}{q_2-q_1}, \\
s_4(q_2)&=q_2 -\frac{\alpha_4}{p_2},\;
s_5(p_2)=p_2-\frac{\alpha_5}{z-q_2} ,\;
s_6(q_1)=q_1+\frac{\alpha_6}{z-p_1-p_2}, \;  
s_6(q_2)=q_2+\frac{\alpha_6}{z-p_1-p_2},
\end{split}
\label{N6si}
\end{equation}
where we only listed those transformations that are not identities and
each $s_i$ is accompanied by transformation \eqref{sialphaj} of
$\alpha_i$. The automorphism $\pi$ acts in this setting
as follows: 
\[ 
\pi :  q_1 \to z-p_1-p_2, \, p_1  \to q_1,\, q_2 \to z-p_2, \,
p_2 \to q_2-q_1,\, \alpha_i \to \alpha_{i-1}\,.
\]
The first-order polynomial solutions \eqref{N6polynomialsols1}-\eqref{N6polynomialsols5}
are expressed in terms of variables defined in relation \eqref{qipij}
as the following solutions to Hamilton equations  \eqref{q1p2eqs} :
\begin{alignat}{3}
q_1&=p_1=p_2=\frac{z}{3},\; q_2=\frac{2z}{3},\; &\quad 
  \alpha_i&=(\mathsf{a},\frac23-\mathsf{a},\mathsf{a}, \frac23-\mathsf{a}
,\mathsf{a}, \frac23-\mathsf{a}) \, ,\label{N6polynomialsols1pq}\\
q_1&=p_1=p_2=\frac{z}{2},\; q_2=z, \;&\quad 
\alpha_i&=(\mathsf{a},1-\mathsf{a},\mathsf{a}, 1-\mathsf{a}
, 0,0)\, ,\label{N6polynomialsols2pq}\\
q_1&=p_1=z,\; q_2=2z,\;p_2=0 \;&\quad 
\alpha_i&=(\mathsf{a},2-\mathsf{a}, \mathsf{a}
,0,-\mathsf{a},0)\, ,\label{N6polynomialsols3pq}\\
q_1&=q_2=p_1=z,\; p_2=0 \;&\quad 
\alpha_i&=(2-\mathsf{a},\mathsf{a}, 0
,0,0,0)\, ,\label{N6polynomialsols4pq}\\
q_1&=q_2=p_2=z,\; p_1=0 \;&\quad 
\alpha_i&=(2-\mathsf{a}, 0,0,\mathsf{a},0,0)\, ,
\label{N6polynomialsols5pq}
\end{alignat}
We notice that the solution \eqref{N6polynomialsols1pq} is a fixed point 
of $\pi^2$ automorphism as it is obvious comparing with its form in 
expression \eqref{N6polynomialsols1}. 

\subsection{Connection of $N=4$ formalism to Painlev\'e V equation }
\label{subsection:H2pain}
It is well-known that equations \eqref{N4feqs} or \eqref{hameqs} 
lead to Painlev\'e V equation. We will here establish this relation  explicitly
in order to relate the parameters of both theories. We first define 
$w = {q}/{z}$. 
Taking a derivative of the top equation in  \eqref{hameqs} and
eliminating $p_z$ and $p$ we obtain the second order equation 
\begin{equation} 
w_{zz}= -\frac{w_z}{z}+\left( \frac{1}{2w}+ \frac{1}{2(w-1)}\right)
w_z^2 + \frac{\alpha w}{z^2 (w-1)}+\frac{\beta(w-1)}{z^2 w} +\gamma w
(w-1) +\delta z^2 w (w-1) (2w-1)\, , 
\label{PVa}
\end{equation}
with 
\begin{equation} 
\alpha=- \frac12 \alpha_3^2,\; \beta= \frac12 \alpha_1^2,\;
\gamma= 2-2 \alpha_2-\alpha_1-\alpha_3=\alpha_4-\alpha_2,\;
\delta=  \frac12 \, .
\label{paramPV}
\end{equation}
We need two additional steps to cast equation \eqref{PVa} into a standard form of
Painlev\'e V equation.

First we perform a change of variables $z \to t$ where 
$t = \epsilon z^2/2$
then followed by  a transformation $y={w}/(w-1)$.

In terms of $y$ equation \eqref{PVa} takes a form of standard Painlev\'e V equation 
\begin{equation} 
y_{t t}
= -\frac{y_t}{t}+\left( \frac{1}{2y}+ \frac{1}{y-1}\right)
y_{t}^2 + \frac{(y-1)^2}{t^2} \left({\bar \alpha} y + {\bar \beta}
\frac{1}{y} \right) + \frac{{\bar \gamma}}{ x} y + {\bar \delta} 
\frac{y (y+1)}{y-1}\, ,
\label{finalPV}
\end{equation}
where 
\begin{equation} 
{\bar \alpha}= \frac18 \alpha_3^2,\;\; {\bar \beta}=- \frac18 \alpha_1^2,\;\;
{\bar \gamma}= - \frac{1}{2 \epsilon }
( 2-2 \alpha_2-\alpha_1-\alpha_3)= \frac{\alpha_2-\alpha_4}{2 \epsilon },\;\;
{\bar \delta}=   -\frac12 \frac{1}{\epsilon^2}\, .
\label{paramPVbar}
\end{equation}
For ${\bar \delta}$ to take a conventional value of $-\frac12$ we need 
$ \epsilon^2 = 1$.

\subsection{Riccati solutions of equations  \eqref{N4feqs}}
\label{subsection:riccati}
Let us reduce equations \eqref{N4feqs} by setting 
either $\alpha_2=0,\; f_2=0, \; f_4=z$ or $\alpha_3=0,\; f_3=0,\;f_1=z$.
Using that $f_3=z-f_1$ in the first case and $f_4=z-f_2$ in the second
case we can rewrite the remaining equations for 
$F_i=f_i/z, i=1,2$ as
\begin{equation}
 \frac{d}{d z} F_i =- z F_i (1-F_i) - \frac{\alpha_i+\alpha_{i+2}}{z}
 F_i +  \frac{\alpha_i}{z}, \quad i=1,2 \, ,
\label{ric1}
\end{equation}
in which we recognize Riccati equations \cite{masuda}. Without losing
generality we will discuss the
solution for the case of $i=1$ with the principal solution given in terms of 
Whittaker functions as
\begin{equation}
F_{1} (z) = -\alpha_1\frac{{\rm WhittakerM}(-\frac14 \alpha_3+\frac14 \alpha_1+1,
-\frac12+\frac14 \alpha_1+\frac14 \alpha_3, \frac12 z^2)}
{z^2 {\rm WhittakerM}(-\frac14 \alpha_3+\frac14 \alpha_1, 
-\frac12+\frac14 \alpha_1+\frac14 \alpha_3, \frac12
z^2)}+\frac{\alpha_1}{z^2}\, .
\label{fwhittaker}
\end{equation}
The above expression becomes a rational function for at least one of the
two parameters $\alpha_1,\alpha_3$ being equal to a negative even integer, and
the other equal to an arbitrary integer but not equal to the opposite of that 
 negative even integer ($\alpha_1+\alpha_3\ne 0$) :
\[
\alpha_i = - 2 n , \qquad \alpha_{i+2}= m \ne 2 n, \;  i=1,3 \;\quad   n\in
\mathbb{Z}_{+}, 0,\; m \in \mathbb{Z}
\, .
\]
For the special case $\alpha_1=0=\alpha_3$  it holds that $F_1=0$.
With the above conditions being satisfied the rational solutions occur for 
Painlev\'e parameters:
\[
{\bar \alpha}= \frac12 n^2, \; {\bar \beta}= - \frac12 (\frac{m}{2})^2
, \;\,
\text{ or} \quad
{\bar \alpha}= \frac12 (\frac{m}{2})^2, \; {\bar \beta}= -\frac12 n^2\, .
\]
Let us recall that since $\alpha_2=0$ then  $\epsilon{\bar  \gamma}= -\alpha_4/2=
-(2-\alpha_1-\alpha_3)/2$ . 
Thus if $\alpha_1=-2n, n\in \mathbb{Z}_{+} $ then we can 
rewrite $\alpha_3$ as $\alpha_3=2 (1+n+\epsilon {\bar \gamma})$.
If $\alpha_3=-2n, n\in \mathbb{Z}_{+}$ then $\alpha_1=2 (1+n+\epsilon \,{\bar \gamma})$.

Riccati equation \eqref{ric1} takes a more familiar look
when we rewrite it in terms of a variable $x= -z^2/2$ :
\[
\frac{d}{d x} F_i= F_i  (1-F_i) - \frac{\alpha_i+\alpha_{i+2}}{2x} F_i +
 \frac{\alpha_1}{2x}\, .
\]
To linearize this equation we set $F_i = w_{i\, x}/w_i $ and
for brevity introduce coefficients 
$b_i= (\alpha_i+\alpha_{i+2})/2$
and $a_i=\alpha_i/2$. In this way we obtain the second-order Kummer's
equation:
\begin{equation}
x w_{i \,xx}+(b_i -x) w_{i \,x} - a_i w_i=0\, .
\label{kummereq}
\end{equation}
We look for solutions of Kummer's equation denoted as $U (a,b,x)$ 
that are polynomials in $x$ of a finite, let us say $n$, degree. This occurs
for $a=-n$ and for $a-b=-n-1$ for $n=0,1,2,3{\ldots} $ and in the latter
case it holds that \cite{nist} :
\begin{equation}
U(a, a +n+1, x) = x^{-a} \sum_{r=0}^{n} \, \binom{n}{r} \, (a)_{r}\,
x^{-r},
\label{Uaan}
\end{equation}
where $(a)_{r}$ is a Porchhammer symbol. 

We will connect this polynomial with the case of $\alpha_3=0$ and 
$a= \alpha_2/2, b=(\alpha_2+\alpha_4)/2$ for $\alpha_i =( \alpha_1 +2
n, -2n,0 , 2-\alpha_1)$,  which we will revisit later in equation \eqref{pochhammer}
in subsection \ref{subsection:item3}, where it will be obtained by an 
action of $T_2^{-n} $ shift operator on polynomial solutions \eqref{solution1}.
For such values of $a$ and $b$ we will need to calculate 
$U(-n, 1-n-\frac{\alpha_1}{2}, x)$. 
Thanks to Kummer's transformation $U(a,b,x)=x^{1-b} U( a-b+1,2-b,x)$
\cite{nist} we obtain a relation
\begin{equation}
U(-n, 1-n-\alpha_1/2, x) = x^{n+\alpha_1/2}\, U (\frac{\alpha_1}{2},
+\frac{\alpha_1}{2}+n+1 ,x)\, ,
\label{Umn}
\end{equation}
which is a polynomial of degree $n$  according to equation \eqref{Uaan}.

For the case of $\alpha_2=0$ we have $a= \alpha_1/2$ and  $b =
\alpha_1/2 +\alpha_3/2$. We will consider $\alpha_i= (\alpha_1, 0 , -2
n , 2-\alpha_1+2n)$, which as shown in subsection \ref{subsection:item3}
are obtained by action of  $T_4^{n} $ shift operator on 
the polynomial solution \eqref{solution1}.
Accordingly, we are dealing with the Kummer function
$ U (\alpha_1/2, \alpha_1/2-n,  x)$. This expression is not a
polynomial as we can verify by explicitly calculating  this function for $n=1$
obtaining $ U (\alpha_1/2, \alpha_1/2-1,  x)= (2 x +\alpha_1-2) e^x$
with $U_x/U$ being however a rational function.  In subsection \ref{subsection:item3}
we will prove that the action of $T_4^{n} $ shift operator on 
the polynomial solution \eqref{solution1} generates solutions of the
Riccati equation \eqref{ric1} for $\alpha_i= (\alpha_1, 0 , -2
n , 2-\alpha_1+2n)$ .

\subsection{Power series representation of $p$ and $q$ variables }
\label{subsection:kova} 
For $N=4$  we will show that $q=j_1+j_2,p=j_2+j_3$ can be represented 
by power series in odd powers  of $z$  and the results are
(up to an action with $\pi$ automorphism and its powers)
\[ q= \sum_{i=1}^2 (c_i z+ e_i z^3 +{\ldots}) , \;\quad 
p= \sum_{i=1}^2 (c_i z +e_i z^3 +{\ldots})\, ,
\]
or
\[ q= \sum_{i=1}^2 (c_i z+e_i z^3 +{\ldots}) , \; \quad 
p= \frac{\alpha_3-\alpha_1}{z} +\sum_{i=1}^2 (c_i z+ e_i z^3 +{\ldots})\, .
\]
The second case can be transformed by  $s_1$ B\"acklund
transformation to the previous case.

Consider power series expansion $j_i= k_i z^{-m}+{\ldots}$ with the first term being lowest
power in $z$ .  Comparing both sides of  equations \eqref{dressingeqseven}
we notice that the lowest terms on the left and the right sides
will be of the order
\begin{equation} z^{-m-1} \sim z^{-2m} + z^{-m-1} (\Psi_{(-2m)} z^{-2m}+{\ldots}
+\Psi_{(0)}) \, ,\label{power}
\end{equation} 
where we use the expansion of 
$\Psi$ in \eqref{psidef} in  powers of $z$ :
\[ 
\Psi = {\ldots} + \frac{\Psi_{(-2)}}{z^2} + \frac{\Psi_{(-1)}}{z}+ 
\Psi_{(0)}+ \Psi_{(+1)} z^1 + {\ldots} \, .
\]
For the terms on both sides of \eqref{power} to match and cancel each other 
we need to take $m=1$ and set all 
$\Psi_{(k)}=0, k<0$. In such case only $\Psi_{0}$ contributes to the above
equation. 

Without losing generality we therefore adopt the expansion 
\begin{equation} j_i(z) =\frac{a_i}{z}+ b_i +c_i z +d_iz^2  +e_i z^3 +
{\ldots} \, .
\label{jpole}
\end{equation}
For expansion in \eqref{jpole} it follows that
\begin{equation} \begin{split}
\Psi_{(-2)} &= a_1^2+a_3^2-a_2^2-a_4^2=-2 (a_1+a_2)(a_2+a_3)\, ,\\
\Psi_{(-1)} &= 2 (a_1 b_1+a_3b_3-a_2 b_2-a_4b_4)=-2 \left( (a_1+a_2)
(b_2+b_3)+(a_2+a_3)(b_1+b_2) \right)\, ,
\end{split}
\label{psim2m1}
\end{equation} 
after we used that $a_4=-a_1-a_2-a_3$ and $b_4=-b_1-b_2-b_3$.

Next we will effectively work with the dressing equations \eqref{dressingeqs}
without $\Psi$ 
to see whether solutions for $j_i=a_i/z +b_i + c_i z$ 
will be such that the divergent terms can be absorbed in $\Psi$
of equation \eqref{dressingeqseven}
via transformation  \eqref{jbardef}:
\[ j_i  \; \to j_i +(-1)^i \frac{1}{2z} \Psi= j_i +(-1)^i \frac{1}{2z} \Psi_{(0)}+
(-1)^i \frac{1}{2z} \Psi_{(1)} z +{\ldots} \, .
\]
On the $z^{-2}$
level of such dressing equations one finds the following expressions:
\begin{equation}  
-(a_i+a_{i+1})=a_{i+1}^2-a_i^2=(a_{i+1}+a_i)(a_{i+1}-a_i), \; i=1,
{\ldots} ,N\, ,
\label{dressm2}
\end{equation}
which imposes that 
\[
a_i+a_{i+1}=0 \;\; \text{or}\;\;a_{i+1}-a_{i}=-1\, ,
\]
for each $i=1,2,3,4$.
There are two independent solutions of the above 
equations:
\begin{align}  
a_i &= ( 1,-1,1,-1)\, a\, , \label{aisols1}\\
a_i &=(a, -a, -1-a, 1+a) \, ,\label{aisols2}
\end{align}
that all satisfy $\sum_{i} a_i=0$. There are other similar solutions that
one can obtain from \eqref{aisols2} 
by acting with $\pi, \pi^2, \pi^3$ transformations to obtain other
solutions like e.g. $ a_i=(a,a-1,1-a,-a)$ and
$a_i=(-1+a,1-a,-a,a)$. 
It therefore suffices to use below the solution \eqref{aisols2}.
The top equation \eqref{aisols1} is such that $a_i+a_{i+1}=0$ for every 
$i=1,2,3,4$. Such divergence can be absorbed by the transformation
\eqref{jbardef} with $\Psi= 2a$. In addition the divergent terms will be absent from
expressions for $p$ and $q$.

The other solution \eqref{aisols2} is such that either
$a_1+a_2=0$ or $a_2+a_3=0$ ensuring $\Psi_{-2}=0$ according to relation
\eqref{psim2m1}. However the divergent terms are such that they can
not be removed the transformation
\eqref{jbardef} and the divergent terms will be present in expressions for 
$p$. Let us illustrate this by applying the transformation 
\eqref{jbardef} with $\Psi=-2(1+a)$. This results in 
  $a_i =(1+2a, -(1+2a), 0, 0)$. 
As we will show below  such divergent terms can be removed by 
a B\"acklund transformation. The calculations done for $N=4$ and $N=6$
suggest that this is  a general feature for all $N$.

Now for solution \eqref{aisols1} we obtain that the condition 
\eqref{psim2m1} for $\Psi_{(-1)}=-2 ((a_1+a_2)
(b_2+b_3)+(a_2+a_3)(b_1+b_2))=0$ is satisfied automatically
and accordingly $b_i$ can be chosen arbitrarily.
For \eqref{aisols2} and the other configurations that can be obtained
from \eqref{aisols2} by $\pi$, we obtain 
conditions $(-1-2a)(b_1+b_2)=0$, $(-1+2a)(b_2+b_3)=0$
and $(1-2a)(b_1+b_2)=0$.
Accordingly $b_i$ can  be chosen arbitrarily if 
$a= \pm 1/2$ or we will have $b_2=-b_3$ or $b_2=-b_1$ condition 
imposing one condition on $b_i$.

Consider now $z^{-1}$ level of the 
equations \eqref{dressingeqseven}
without $\Psi$. With such redefined system one obtains on the $z^{-1}$
level $0=a_{i+1} b_{i+1} - a_i b_i$.
For the solutions in \eqref{aisols1} and \eqref{aisols2}
we find that we can write $b_i=b (1,-1,1,-1)$ and we can set $b=0$
without losing any generality as the terms can be added or removed by the transformation
\eqref{jbardef}. Similar conclusion
can be obtained for other coefficients of terms with $z$  to the even
power: $z^{2k}$. Such terms will not contribute to $q=j_1+j_2, p=j_2+j_3$
and we don't need to consider them in what follows.

Consider now $z^{0}$  levels of the 
equations \eqref{dressingeqs} :
\begin{equation}
(1+2 a_i ) c_i+(1- 2 a_{i+1} ) c_{i+1} =  \alpha_i , \quad
i=1,{\ldots} , N\, ,
\label{ciz0}\\
\end{equation}
using that $b_i^2=b^2_{i+1}$.

We first plug values for $a_i$ from \eqref{aisols1} into
the above equation to obtain 
\[
a= \frac12 (-1+\alpha_1+\alpha_3)\, ,
\]
using that $\sum_i c_i=1$.
For $a_i$ given in \eqref{aisols2} we find
\begin{equation}
a= \frac12 (-1+\alpha_1-\alpha_3)\, ,
\label{aaa}
\end{equation}
and 
\begin{equation}
c_1+c_2 = \frac{\alpha_1}{\alpha_1-\alpha_3}\, .
\label{c12}
\end{equation}
We will now apply our results to $q=j_1+j_2, p=j_2+j_3$ variables.
For $a_i =(a, -a, -1-a, 1+a)$ and $a$ given in \eqref{aaa}
it holds that 
$-(1+2a)=\alpha_3-\alpha_1$ and
\begin{equation}
q= c_{12} z + e_{12}z^3 +{\ldots} , \;\;\quad 
p= \frac{\alpha_3-\alpha_1}{ z} + c_{23} z+ e_{23} z^3+{\ldots} \, .
\label{azp}
\end{equation}
Here for
brevity we introduced $c_{12}=c_1+c_2$ given in equation \eqref{c12}.
Explicit calculation gives 
\[ 
c_{23}=\frac{\alpha_3^2+\alpha_3\alpha_2-2
\alpha_3+\alpha_1^2+\alpha_1^2+\alpha_2\alpha_1-2 \alpha_2-2
\alpha_1}{\alpha_3^2 -2 \alpha_3\alpha-1-4+\alpha_1^2}\, .
\]
It follows that the singular term in $p$ in \eqref{azp} can be removed by $s_1$
transformation : $q \to q$, $p \to p+\alpha_1/q$ with 
\begin{equation}
\begin{split}
\frac{\alpha_1}{q}&= 
\frac{1}{\frac{1}{\alpha_1} ( - \frac{\alpha_1}{\alpha_3-\alpha_1} z + e_{12}z^3
+{\ldots} )}= \frac{1}{\frac{-z}{\alpha_3-\alpha_1}(1 -\frac{z^2
e_{12}(\alpha_3-\alpha_1)}{\alpha_1}+{\ldots} )}\, ,\\
&=-\frac{\alpha_3-\alpha_1}{z}(1+ \frac{z^2
e_{12}(\alpha_3-\alpha_1)}{\alpha_1} + z^4{\ldots} )
= -\frac{\alpha_3-\alpha_1}{z} -z 
\frac{e_{12} (\alpha_3-\alpha_1)}{\alpha_1}
+z^3{\ldots} )\, ,
\end{split}
\label{alpha1q}
\end{equation}
which shows that the transformed $p$ given by $ p+\alpha_1/q$ will no
longer contain a singular term. Its power expansion will start with the term
proportional to $z$ and will only contain odd powers of $z$.

The initial position of the pole can be obviously moved from $p$ to
$q$ by the $\pi$ automorphism. This will lead to $s_1$ being
transformed by  $\pi$ to other $s_i$, which will remove the divergent terms.
With this understanding we continue to
consider the above configuration without any loss of generality.
One can therefore effectively only consider the case of 
$a_i =a (1, -1, 1, -1)$ from \eqref{aisols1} with
\[ q= c_{12} z + e_{12}z^3 +{\ldots} , \;\;
p=  c_{23} z+ e_{23} z^3+{\ldots} \, ,
\]
with 
\[ c_{12}=  \frac{\alpha_1}{\alpha_1+\alpha_3}, \;
c_{23}=\frac{\alpha_2}{2-\alpha_1 - \alpha_3}\, .
\]
Amazingly the first terms of a general expression for $q,p$ agree
with a general formula
\begin{equation}
q=\frac{\alpha_1}{\alpha_1+\alpha_3} z ,\;\;
p=\frac{\alpha_2}{2-\alpha_1-\alpha_3} z, \;\;
\label{genwata}
\end{equation}
that reproduces all the cases of \eqref{wata3}-\eqref{wata6} for
the corresponding values of $\alpha_i$.

Let us illustrate all this by the following example.
\begin{exmp}
The solution
\begin{equation}
q= \frac{z(-468 +z^4)}{2(z^4-324)},
\;\; p= \frac{z^2-18}{2z},\;
\alpha_1=\frac{13}{2},\alpha_2=-1, \alpha_3=-\frac{5}{2} \, ,
\label{oursol2example}
\end{equation}
is taken from reference \cite{ullate}, where it was obtained using Maya
diagram techniques. 
Clearly $p=z/2-9/z$ contains a singularity. Note that indeed
$-9/z=(\alpha_3-\alpha_1)/ z$ in agreement with relation  \eqref{azp}.
Applying $s_1$ we get:
\begin{equation}
q= \frac{z(-468 +z^4)}{2(z^4-324)},
\;\; p= \frac{z(z^4+8z^2-468)}{2(z^4-468)},\;
\alpha_1=-\frac{13}{2},\alpha_2=\frac{11}{2}, \alpha_3=-\frac{5}{2} \, ,
\label{s1oursol2}
\end{equation}
with polynomial expansions:
\[ q(z)= \frac{13}{18} z +\frac{1}{1458} z^5+\frac{1}{472392} z^9
+{\ldots} ,\;\; p(z)= \frac{1}{2} z-\frac{1}{117}
z^3-\frac{1}{54756}z^7 + {\ldots} \, .
\]
Note that $13/18=\alpha_1/(\alpha_1+\alpha_3)$. We will show below 
how to derive the rational solutions  \eqref{s1oursol2}
from the seed solutions \eqref{solution2}-\eqref{solution1} (or 
\eqref{wata3}-\eqref{wata6}) by the shift operators.
\end{exmp}

Applying equations \eqref{dressm2} and \eqref{ciz0} to $N=6$
we find that the number of solutions increased from two to three 
 (up to an action of $\pi$ automorphism) and they are given by:
\begin{align}
a_i &= a\, ( 1,-1,1,-1,1,-1),\qquad \qquad \qquad \quad \; 
a=-\frac12 (1-\alpha_1-\alpha_3-\alpha_5) \, ,\label{aisolN6-1}\\
a_i &=(a, -a, -1-a, 1+a,-1-a,1+a), \quad 
a=-\frac12 (1+\alpha_1-\alpha_3-\alpha_5)  \, ,\label{aisolN6-2}\\
a_i &=(a, -a, -1-a, 1+a,a,-a), \qquad \quad\quad \;\; a=-\frac12
(1+\alpha_1-\alpha_3+\alpha_5) \label{aisolN6-3}\, ,
\end{align}
for expansions $j_i (z) = a_i /z +c_i  z + {\ldots} , i=1,{\ldots} ,6$.
For solutions \eqref{aisolN6-2} and \eqref{aisolN6-3} there will be
poles in expansions of $p_i, i=1,2$.

Note that from equations \eqref{ciz0} we find $c_1+c_2=\alpha_1/(1+2
a)$ and $c_3+c_4=-\alpha_3/(1+2a)$ where $a$ is given in relations
\eqref{aisolN6-2} and \eqref{aisolN6-3}, respectively.

In case of solution \eqref{aisolN6-2} the expansion of $p_1$ starts with a
pole $p_1 = -(\alpha_1-\alpha_3-\alpha_5) /z + {\ldots} $ while the
expansion of $q_1$ is $q_1=(c_1+c_2)z +{\ldots} =
\alpha_1 z /(\alpha_1-\alpha_3-\alpha_5) +{\ldots} $.
Consequently the action of $s_1$ on $p_1$ removes the pole similarly
to what we have seen for $N=4$ case in expression \eqref{alpha1q}.

In case of solution \eqref{aisolN6-3} both expansions of $p_i, i=1,2$
will start with divergent terms: $p_1= -(\alpha_1-\alpha_3+\alpha_5) /z + {\ldots}$
and $p_2= (\alpha_1-\alpha_3+\alpha_5) /z + {\ldots}$.
Since $q_1= (c_1+c_2)z +{\ldots}$ and $q_2= (c_1+c_2+c_3+c_4)z +{\ldots}$
we easily find that $q_1-q_2= \alpha_3 z/(\alpha_1-\alpha_3+\alpha_5)+{\ldots} $.
Consequently the action of $s_3$ from equation \eqref{N6si}
on $p_1$ and $p_2$ will remove these divergencies. For those solutions
that are obtained from solutions \eqref{aisolN6-2} or \eqref
{aisolN6-3} by acting with automorphism $\pi$ or its powers the
divergencies will be removed by appropriate B\"acklund transformations
that are conjugations of $s_1, s_3$, e.g. $\pi s_1 \pi^{-1}$, $\pi s_3
\pi^{-1}$ etc.

\section{Construction of Rational Solutions}
\label{section:rational}
In this section, we will describe a method to derive  all rational solutions that are
obtainable from the first-order polynomial solutions 
of dressing equations \eqref{dressingeqseven}
via combined actions of fundamental shift operators $T_i, i=1,{\ldots}
,N$ from \eqref{shifts}. 
\subsection{Summary of the results for $\pmb{N=4}$}
For $N=4$ the seeds solutions \eqref{solution2} and \eqref{solution1} 
 of dressing equations \eqref{dressingeqseven} are equivalent to Watanabe's algebraic solutions
\eqref{wata3}-\eqref{wata6} in the setting of Hamilton equations \eqref{hameqs}.
It is convenient to give the classification of solutions in terms of parameters
$\alpha_1,\alpha_3, (\alpha_2-\alpha_4)/2$ of the dressing chain
equations that define the Painlev\'e V
parameters ${\bar \alpha}, {\bar \beta}, {\bar \gamma}$ via relations
\eqref{paramPVbar} with ${\bar \delta}$ parameter 
being non-zero  and here equal to  
${\bar \delta} =-1/2$ (for $\epsilon^2=1$).

\label{page:items}
The rational solutions obtained by acting with the shift operators 
fall into three classes of  parameters
$\alpha_1,\alpha_3, (\alpha_2-\alpha_4)/2$ and  ${\bar \alpha}, {\bar \beta}, {\bar \gamma}$ 
depending on whether the fundamental 
shift operators act
on solutions 
\begin{itemize}
\item $j_i=(z/4) (1,1,1,1)$ from \eqref{solution2} (items
(Ia,Ib) and item (II)). In case of item (II) an intermediary step
of acting with $s_1$ in addition to the shift operators is involved, 
see f.i. equation \eqref{azp}.  
\item $j_i=(z/2)(1,1,-1,1)$ from \eqref{solution1}
(items (IIIa,IIIb)). 
\end{itemize}
These three cases are as follows:

\begin{enumerate}
\item[(I)]
\[
{\alpha}_1 = A +2 n_1-2 n_2,\; 
\alpha_3= A +2 n_3-2 n_4,\; 
\frac{{\alpha}_2-{\alpha}_4}{2}=n_2-n_3-n_4+n_1, 
\]
with $n_i \in  \mathbb{Z}, i=1,{\ldots} ,4$ and $A$ arbitrary. The above implies either (Ia) or (Ib): 
\item[(Ia)]
$ {\bar \alpha} = \frac12 (a)^2$, \, $ {\bar \beta} =- \frac12
(a+n)^2 $ and  $ {\bar \gamma} =\epsilon m$ 
where $m+n$ is even and equal to $2(n_1-n_3)$ and $a=A/2+n_3-n_4$ arbitrary,
\item[(Ib)]
$ {\bar \alpha} = \frac12 (b+n)^2$, \, $ {\bar \beta} =- \frac12
(b)^2 $ and  ${\bar \gamma} =\epsilon m$ 
where $m+n$ is even and equal to $2(n_2-n_4)$ and $b=A/2+n_1-n_2$ arbitrary
\item[(II)]
\[
\begin{split}
	{\alpha}_1 &=1+2 n_1-2 n_2,\;{\alpha}_2= - A +2 n_2-2 n_3,\\
	{\alpha}_3&= 1+2 n_3-2 n_4,\quad {\alpha}_4=  A +2
	n_4-2 n_1\, ,
\end{split}
\]
which imply
\[ {\bar \alpha}= \frac12  \left( \frac{1}{2}+m\right)^2, \;
{\bar \beta}=- \frac12 \left( \frac{1}{2}+n \right)^2,  \;
{\bar \gamma}=(-A +n+m) \epsilon \, ,
\]
where $A$ is arbitrary
and $n,m$ are integers.

\item[(IIIa)]
\[
{\alpha}_1 = A +2 n_1+2n_2,\; 
{\alpha}_3= -2 n_4 ,\; 
\frac{{\alpha}_2-{\alpha}_4}{2}=-\frac{{\alpha}_4}{2}=
\frac{A}{2} -1 -n_4+n_1-n_2, \;  \;\;
n_2, n_4  \in  \mathbb{Z}_{+},\; n_1 \in  \mathbb{Z}\, ,
\]
with $A$ arbitrary and $\mathbb{Z}_{+}$ that includes positive integers and zero. Accordingly, eliminating the arbitrary number $A$ from the above equations we can write 
\[
{\bar \alpha}= \frac18 \alpha_3^2 = \frac12  \left( n\right)^2, \;\;
{\bar \beta}=-\frac18 \alpha_1^2= - \frac12 \left(\epsilon {\bar \gamma} +1 + m \right)^2,  \;\;
\]
where $n=n_4 ,m=n_4+2n_2  \in \mathbb{Z}_{+}$ and with  $n+m$ being an even integer.

\item[(IIIb)] 
\[{\alpha}_1 =-2 n_2,\; { \alpha}_3=A+ 2 n_3+2 n_4,\quad 
 \frac{{\alpha}_2-{\alpha}_4}{2}= 1-\frac{A}{2}+ n_2- n_3+n_4,
 \;\;  n_2, n_4 \in \mathbb{Z}_{+},\; n_3 \in \mathbb{Z}\, ,
\]
with $A$ arbitrary. $\mathbb{Z}_{+}$ 
includes positive integers and zero. Accordingly, eliminating the 
arbitrary number $A$ from the above equations we can write 
\[
{\bar \alpha}= \frac18 \alpha_3^2=\frac12  \left( -\epsilon {\bar \gamma}+1+m\right)^2, \;\;
{\bar \beta}=-\frac18 \alpha_1^2=- \frac12 \left(  n \right)^2,  \;\;
\]
where $n=n_2,m=n_2+2n_4  \in \mathbb{Z}_{+}$ and with $n+m$ being an even integer.
\end{enumerate}
\label{page:items1}

Comments : Integers $n ,m $ in (IIIa) and (IIIb) have been derived as
positive integers. However they both enter quadratic expressions in
which their overall sign can be reversed.

\subsection{Applying the shift operators to obtain rational solutions}
\label{subsection:shiftrational}
For $N=4$ we will show how to reproduce 
items (I)-(III) listed on the page \pageref{page:items}
in the setting of  Painlev\'e V equation using the following
construction :
\begin{itemize}
\item  The seeds of  all rational solutions 
are the  first-order  
polynomial solutions   \eqref{solution2}, \eqref{solution1} and its $\pi$ variants. Note that these
seeds solutions all depend on an arbitrary real parameter customary chosen 
here as $\mathsf{a}$.
\item A class of rational solutions that 
can be obtained by  successive operation by shift operators $T_i$,
defined in the next subsection \ref{subsection:shiftops}, of the form :
\begin{equation}
T_1^{n_1}T_2^{n_2}T_3^{n_3}T_4^{n_4} , \qquad n_i \in  \mathbb{Z}\, ,
\label{T1234}
\end{equation}
on polynomial solutions \eqref{solution2} can be expanded in 
positive power series in $z$ and do not contain a pole singularity and if
necessary (like in case of equation \eqref{azp}) having this  singularity removed by $s_1$ B\"acklund
transformation. These two cases are described by the  parameters presented
in the above items I and II, respectively. 
\item A class of rational solutions 
obtained  from the seeds polynomial solutions \eqref{solution1} 
will be derived by  successive operation with shift operators $T_i$
of the type
\begin{equation}
T_i^{n_i} T_j^{n_j} T_k^{-n_k}, \qquad     n_j,n_k \in  \mathbb{Z}_{+},
\; n_i  \in  \mathbb{Z}\, ,
\label{T1234restricted}
\end{equation}
for distinct $i,j,k$ and $ \mathbb{Z}_{+}$ that contains positive
integers and zero as only actions with shift operators given in
equation \eqref{T1234restricted} that are not causing divergencies.
The results are summarized in the item III on page
\pageref{page:items1}.
\end{itemize}
We conclude that the well-known  fundamental results
 on classification of rational solutions of Painlev\'e V equation
first  presented in \cite{kitaev}  
are here obtained by acting with the operators \eqref{T1234}
on the first-order polynomial solutions \eqref{solution2} and \eqref{solution1}. In the latter case we will encounter 
restrictions on those values of $n_i$ for which the operators \eqref{T1234} 
are well-defined, as indicated in equation \eqref{T1234restricted}.
See also \cite{ohta} that derived rational solutions 
described above in items (Ia,Ib) and (II)  
via shift operators
acting on solutions expressed by $\tau$ functions and
corresponding to  \eqref{solution2}. The results
of reference \cite{kitaev} were summarized succinctly in
\cite{gromak}.

\subsection{The fundamental shift  operators for $\pmb{A_{N-1}^{(1)}}$ }
\label{subsection:shiftops}

To analyze transformations under the shift operators which we will
introduce in this subsection it is convenient
to first introduce  the following representation of $\alpha_i$ parameters for
$N=4$ case :
\begin{equation}
\alpha_1= 2(v_2-v_1),\;\;\; \alpha_2 = 2 (v_3-v_2) , \;\;\;
\alpha_3= 2(v_4-v_3),\;\;\; \alpha_4= 2+2(v_1-v_4)\, .
\label{alphav}
\end{equation}
One checks that 
\[
\alpha_1+\alpha_2+\alpha_3+\alpha_4= 2
+2(v_2-v_1+v_3-v_2+v_4-v_3+v_1-v_4)=2\, ,
\]
is satisfied automatically without imposing any condition on $v$'s.

Obviously adding a constant term to all $v_i$ will
not change the final result in \eqref{alphav} and thus we have an equivalence:
\begin{equation}
(v_1,v_2,v_3,v_4) \sim  (v_1+c ,v_2+c,v_3+c,v_4+c).
\label{equivalence}
\end{equation}

The B\"acklund transformations $s_i, i=1,2,3$ act in terms of $v_i$
simply as permutations between $v_i$ and $v_{i+1}$ :
$s_i:  v_i \leftrightarrow v_{i+1}$, while  $s_i(v_j)=v_j, j\ne i, i+1$. 
The automorphism $\pi$ acts as
follows: 
$\pi(v_i)=v_{i-1}, i=2,3,4$ and $\pi(v_1)=v_{4}-1$.

Next we introduce the shift operators 
\begin{equation}
T_1=\pi s_3s_2s_1, \quad T_2=s_1\pi s_3s_2, \quad T_3=s_2s_1\pi s_3, 
\quad T_4=s_3s_2s_1\pi, 
\label{trans}
\end{equation}
that act as simple translations on the $v_i$ variables:
$T_i (v_j)= v_j-\delta_{i,j}$ leading to:
\begin{equation}
T_i (v_i) = v_i-1, T_i(v_j)=v_j  \longrightarrow
 \begin{cases} T_i(\alpha_i)=\alpha_i+2, \\
T_i(\alpha_{i-1})=\alpha_{i-1}-2\end{cases} \, ,
\label{Tivalpha}
\end{equation}
or 
\begin{equation}
\begin{split}
T_1  (\alpha_1, \alpha_2, \alpha_3, \alpha_4) &=
 (\alpha_1+2, \alpha_2, \alpha_3, \alpha_4-2) \, ,\\
T_2  (\alpha_1, \alpha_2, \alpha_3, \alpha_4) &=
 (\alpha_1-2, \alpha_2+2, \alpha_3, \alpha_4) \, ,\\
T_3  (\alpha_1, \alpha_2, \alpha_3, \alpha_4) &=
 (\alpha_1, \alpha_2-2, \alpha_3+2, \alpha_4) \, , \\
 T_4  (\alpha_1, \alpha_2, \alpha_3, \alpha_4) &=
 (\alpha_1, \alpha_2, \alpha_3-2, \alpha_4+2) \, .
\label{Tialpha}
\end{split}
\end{equation}
Comparing expressions \eqref{Tialpha} and \eqref{Tivalpha} 
we see that in the $v_i$ representation it 
is very convenient to study  how the parameter
space of solutions of dressing equation is being formed under actions
of the shift operators. Generally the orbit of $v_i=(v_1,v_2,v_3,v_4)$
under a action with $T_1^{n_1}T_2^{n_2}T_3^{n_3}T_4^{n_4}$ 
from equation \eqref{T1234} will be described by
$v_i=(v_1-n_1,v_2-n_2,v_3-n_3,v_4-n_4)$. 
We are then able to associate a rational solution to each point of the
orbit following approach of subsection \ref{subsection:shiftrational}.

It is easy to extend the definition of the fundamental shift 
operators  to arbitrary $N$  \cite{witte,ohta,noumi98} :
\begin{equation}
T_1= \pi s_{N-1} \cdots s_2 s_1, \; T_2 = s_1 \pi s_{N-1} \cdots s_2,
{\ldots} , T_N= s_{N-1} \cdots s_2 s_1 \pi\, ,
\label{shifts}
\end{equation}
that for every $N$ generate the weight lattice of $A_{N-1}^{(1)}$.
The shift operators  commute with each other 
\[ T_iT_j = T_j T_i\, ,
\]
and satisfy $T_1T_2 \cdots T_{N}=1$, where we used that $\pi^N=1$ 
and that $\pi s_i= s_{i+1} \pi$.
These operators act on parameters $\alpha_i$ as 
\begin{equation}
T_i(\alpha_{i-1})=\alpha_{i-1}-2, \quad T_i(\alpha_i)=\alpha_i+2, 
\quad T_i(\alpha_j)=\alpha_j~(j \ne i-1,i),
\label{Tialphas}
\end{equation}
and further satisfy $ \pi T_i= T_{i+1} \pi,\;T_i (\Phi)=\Phi,\; T_i (\Psi)=-\Psi$.
The inverse shift operators for $N=4$ are :
\begin{equation}
	T_1^{-1} =s_1s_2s_3 \pi^3, \quad T_2^{-1}=s_2 s_3 \pi^3 s_1, \quad T_3^{-1}=s_3\pi^3 s_1 s_2, 
	\quad T_4^{-1} =\pi^3 s_1s_2s_3\, .
\label{inverseT}
\end{equation}
For convenience we also  list the shift operators for  $N=6$:
\begin{equation}
\begin{split}
T_1&= \pi s_{5} s_4 s_3 s_2 s_1, \; T_2 = s_1 \pi s_{5}s_4 s_3 s_2,\;
 T_3 = s_2s_1 \pi s_{5}s_4 s_3,\; T_4 = s_3 s_2s_1 \pi s_{5}s_4 \\
T_5 &= s_4 s_3 s_2s_1 \pi s_{5},\; 
T_6= s_5 s_4 s_3 s_2 s_1 \pi\, ,
\end{split}
\label{N6shifts}
\end{equation}
and their inverse 
\begin{equation}
\begin{split}
	T_1^{-1} &=s_1s_2s_3s_4 s_5  \pi^{-1}, \;
	T_2^{-1}=s_2 s_3 s_4 s_5 \pi^{-1} s_1, \;
	T_3^{-1}=s_3 s_4 s_5 \pi^{-1} s_1 s_2, 
	\; T_4^{-1} =s_4 s_5 \pi^{-1}  s_1s_2s_3,\\
T_5^{-1} &= s_5 \pi^{-1}  s_1s_2s_3 s_4,\; 
T_6^{-1} =  \pi^{-1}  s_1s_2s_3 s_4s_5 
	\end{split}\, .
\label{inverseT6}
\end{equation}

Within the framework of dressing chain equations with B\"acklund
transformations \eqref{Tinewj} it is actually possible to
establish a general  transformation rules for the shift operator
$T_i$ acting on $j_{i+1}$,  $j_{i+2}, {\ldots} $ for $i=1.,{\ldots} ,N$,
which applies to $N=4,6$ and the initial configurations 
\eqref{solution2}, \eqref{N6polynomialsols1} :
\begin{equation}
\begin{split}
 j_{i+1 ,\, n+1} &=  T_i ( j_{i+1 ,\, n})= j_{i \, n}- 
 \frac{\mathsf{a}+2n}{ j_{i ,\, n}+ j_{i+1 ,\, n}} \\
 j_{i+2 ,\, n+1} &=  T_i ( j_{i+2 ,\, n})= j_{i+1 ,\, n}+ 
 \frac{\mathsf{a}+2n}{ j_{i ,\, n}+ j_{i+1 ,\, n}}\\
&-\frac{4/N+2n}{j_{i+1 ,\, n}+ j_{i+2 ,\, n}+ j_{i+3 ,\, n}- T_i (
j_{i+1 ,\, n})}\\
j_{i+3 ,\, n+1} &=  T_i ( j_{i+3 ,\, n})= j_{i+2,\, n}
+\frac{4/N+2n}{j_{i ,\, n}+j_{i+1 ,\, n}+j_{i+2 ,\, n} -T_i (j_{i+1
,\, n})}\\
&-\frac{4/N+a+2n}{j_{i ,\, n}+j_{i+1 ,\, n}+j_{i+2 ,\, n}+j_{i+3 ,\, n} -
T_i (j_{i+1,\, n}+j_{i+2,\, n})}
 \end{split}
 \label{N6arbitrary1recurj}
\end{equation}
etc., where $j_{i+k,\, n}= T_i^n (j_{i+k, \, 0})$ with $j_{i+k, \, 0}=z/N$
and $k=1,2 {\ldots} $.
The above equations lead to
\begin{equation}
\begin{split}
j_{i+1 ,\, n+1}+j_{i+2 ,\, n+1}&= 
T_i ( j_{i+1 ,\, n}+j_{i+2 ,\, n})
= j_{i \, n}+j_{i+1 ,\, n}
\\&- (1+\frac{N}{2} n ) \frac{4}{N}   \frac{j_{i ,\, n}+ j_{i+1 ,\,
 n}}{(j_{i ,\, n}+ j_{i+1 ,\, n})
 (j_{i+1 ,\, n}+ j_{i+2 ,\, n})+ \mathsf{a}+2n}
\end{split}
\label{N6arbitraryT1recurp}
\end{equation}
which for  $i=1$ will lead to recurrence relations for $p=j_2+j_3$ in
case of $N=4$ and for $p_1=j_2+j_3$ in
case of $N=6$. These recurrence relations will establish Umemura
polynomial solutions as will be shown below.

\subsection{Shift operators  acting on the solution
$\pmb{j_i=\frac{z}{4} (1,1,1,1)}$ in equation \pmb{\eqref{solution2}} }
\label{subsection:item1&2}

\subsubsection{Parameters of the solutions obtained from the seed
solution 
$\pmb{j_i=\frac{z}{4} (1,1,1,1)}$ by action of the shift operators}

Consider solutions  \eqref{solution2} 
 such that $(\alpha_1, \alpha_2, \alpha_3, \alpha_4) =
(\mathsf{a},1-\mathsf{a},\mathsf{a},1-\mathsf{a})$ with an arbitrary parameter 
$\mathsf{a}$ and $q=p=z/2$.
According to relation \eqref{Tialpha} these solutions under action of 
\eqref{T1234} will have  the following final 
parameters $\alpha_1,{\alpha}_2,
{\alpha}_3, {\alpha}_4$:
\begin{equation}
{\alpha}_1 = \mathsf{a} +2 n_1-2 n_2,\, 
{\alpha}_3= \mathsf{a} +2 n_3-2 n_4,\;
{\alpha}_2= 1-\mathsf{a}+2 n_2-2 n_3,\,
{\alpha}_4= 1-\mathsf{a}+2 n_4-2 n_1\, .
\label{pre-baralphas1}
\end{equation}
Thus in agreement with item I on page \pageref{page:items} we find 
\begin{equation}
\begin{split}
\frac{{\alpha}_1 - {\alpha}_3}{2} 
&= n_1- n_2- n_3+ n_4=k_1-k_2= 2 k_{-} \\
\frac{{\alpha}_2-{\alpha}_4}{2}&= n_1-n_3+n_2-n_4=k_1+k_2=2
k_{+}\, ,
\end{split}
\label{baralphasdikff}
\end{equation}
where we introduced 
\begin{equation}
k_1=n_1 - n_3, \quad k_2=n_2-n_4, \quad k_{\pm}= \frac12 (k_1 \pm k_2)
\, .
\label{kis}
\end{equation}
In terms of these parameters we can decompose 
$T_1^{n_1}T_2^{n_2}T_3^{n_3}T_4^{n_4}$  into a product of different
factors 
\begin{equation}
T_1^{n_1}T_2^{n_2}T_3^{n_3}T_4^{n_4}= T_1^{k_1} T_2^{k_2} (T_1T_3)^{n_3}
(T_2T_4)^{n_4}= (T_1T_2)^{k_{+}} (T_1T_2^{-1})^{k_{-}} (T_1T_3)^{n_3}
(T_2T_4)^{n_4}\, ,
\label{kpmn}
\end{equation}
with each factor acting independently of the others on parameters in
equation \eqref{baralphasdikff}. 
Their
action on expression  \eqref{solution2} with 
$(\mathsf{a},1-\mathsf{a},\mathsf{a},1-\mathsf{a})$
induces the following transformations:
\begin{enumerate}
\item $(T_1T_3)^{n_3}$ increases arbitrary parameter $\mathsf{a}$ : $\mathsf{a}
\to \mathsf{a}+2 n_3$ but leaves $q=p=z/2$ 
of equation  \eqref{wata3} unchanged.
\item $(T_2T_4)^{n_4}$ decreases arbitrary parameter $\mathsf{a}$ : $\mathsf{a}
\to \mathsf{a} - 2 n_4$ but leaves $q=p=z/2$ 
of equation  \eqref{wata3} unchanged.
\item $(T_1T_2)^{k_{+}} $ increases $\frac12 ({\alpha}_2-{
\alpha}_4) \to \frac12 ({\alpha}_2-{\alpha}_4) + 2 k_{+}$
\item  $(T_1T_2^{-1})^{k_{-}}$ increases $\frac12 ({\alpha}_1-{
\alpha}_3) \to \frac12 ({\alpha}_1-{ \alpha}_3) + 2 k_{-}$
\end{enumerate}
The conclusion in point 1. follows 
easily from the transformation rule :
\begin{equation} (T_1 T_3)^k (j_i) =   (T_2 T_4)^{-k} (j_i)=  
\frac{z}{4} + (-1)^{i+1} \frac{2k}{z}
\label{T1T3ji}
\end{equation}
where $j_i=z/4$  is one of the components of solution
\eqref{solution2}. Similar argument applies to point 2. since
$T_1T_2T_3T_4=1$.
The first two top transformations in points, 1. and 2., do not induce
any change in 
$\frac12 ({\alpha}_2-{ \alpha}_4) $ nor in 
$\frac12 ({ \alpha}_1-{ \alpha}_3) $, thus 
the shift operators $(T_1T_3)^{n_3}$ and 
$(T_2T_4)^{n_4}$ increase equally 
Painlev\'e V parameters ${\bar \alpha}$ and ${\bar \beta}$ and are not
changing $\epsilon \gamma$ parameter.
The above discussion shows that the two seed configurations 
$(\mathsf{a},1-\mathsf{a},\mathsf{a},1-\mathsf{a})$ and 
$(\mathsf{b},1-\mathsf{b},\mathsf{b},1-\mathsf{b})$ both
corresponding  to the solution \eqref{wata3}   
with parameters $\mathsf{a}$ and $\mathsf{b}$ such that 
$\mathsf{b}=\mathsf{a}+2 m$,
with $m$ being an integer, can be connected by the transformation
$(T_1T_3)^{n_3}  (T_2T_4)^{n_4}$ with $m=n_3-n_4$,
that leave $q=p=z/2$ of equation  \eqref{wata3} unchanged.
Thus they both can give rise to identical solution $y,(\alpha,\beta,
\gamma, \delta)$ of the Painlev\'e V equation via actions of different 
fundamental shift operators. However this ambiguity 
disappears when the two seed solutions are considered
as solutions \eqref{solution2} of the dressing chain 
since their $j_i (z)$ components will
transform non-trivially under $(T_1T_3)^{n_3}  (T_2T_4)^{n_4}$
according to relation \eqref{T1T3ji} as long as $n_3 \ne n_4$.

The shift operator $(T_1T_2)^{k_{+}} $ increases 
$\epsilon \gamma$ by $2 k_{+}$, 
while $(T_1T_2^{-1})^{k_{-}}$ changes a difference between ${\bar \alpha}$
and ${\bar \beta}$ of Painlev\'e V parameters. To illustrate how 
the Painlev\'e V parameters ${\bar \alpha}, {\bar \beta}, {\bar \gamma}$
transform under the above combinations of shift operators we 
recall expressions \eqref{paramPVbar}
and take into account expressions \eqref{pre-baralphas1} to obtain  :
\begin{equation}
{\bar \alpha}= \frac{\left(\mathsf{a}/2 + n_3- n_4\right)^2}{2}, \,
{\bar \beta}=- \frac{\left(\mathsf{a}/2 + n_1- n_2 \right)^2}{2},  \,
{\bar \gamma}= \epsilon(n_2-n_3-n_4+n_1) \, .
\label{baralphas1a}
\end{equation}
In terms of integers $k_{\pm}$ the above expressions can be rewritten
succinctly as: 
\[ {\bar \alpha}= \frac12  \left( \sqrt{-2 {\bar \beta}}
+n_2-n_1+n_3-n_4\right)^2 = \frac12  \left( \sqrt{-2 {\bar \beta}}
-2 k_{-}\right)^2, \;{\bar \gamma}=2 \epsilon k_{+}\, .
\]

Sometimes one encounters a pole in an initial expression for $p$ like
 it was the
 case in solution \eqref{oursol2example},
 where  $s_1$ was used to remove the pole from $p$.
To cover such case we apply $s_1$ B\"acklund transformation to obtain
a configuration $(-\mathsf{a},1,\mathsf{a},1)$. Then applying $\pi$
automorphism we arrive at
\[ ( 1, -\mathsf{a},1,\mathsf{a})\, .\]
Acting with $T_1^{n_1}T_2^{n_2}T_3^{n_3}T_4^{n_4}$ from \eqref{T1234}
will yield: 
\begin{equation}
\begin{split}
{\alpha}_1 &=1+2 n_1-2 n_2,\quad {\alpha}_2= - \mathsf{a} +2 n_2-2 n_3,\\
 { \alpha}_3&= 1+2 n_3-2 n_4,\quad {\bar \alpha}_4=  +\mathsf{a} +2
n_4-2 n_1,\label{pis1solution2}
\end{split}
\end{equation}
with 
\[
{\bar \alpha}= \frac12  \left( \frac{1}{2}+n_3- n_4\right)^2, \;\;
{\bar \beta}=- \frac12 \left( \frac{1}{2}+n_1- n_2 \right)^2,  \;\;
{\bar \gamma}=\epsilon (-\mathsf{a}  +n_2-n_3-n_4+
n_1), \; 
\]
setting $\mathsf{a}=A, n_3=0,n_4=-m,n_1=n,n_2=0$ we get item (II)
on page \pageref{page:items},
in agreement with  \cite{kitaev}, see also  \cite{ohta}.

\noindent
\begin{exmp} Consider again the case of solution \eqref{oursol2example}
with
\[ \alpha_1=\frac{13}{2},\quad\alpha_2=-1,\quad \alpha_3=-\frac{5}{2}
,\quad
\alpha_4=-1\, ,
\]
and $p=z/2-9/z$ that contains a pole that can be
removed by $s_1$. Fitting the above $\alpha$'s into relation
\eqref{pre-baralphas1} does not work since the method works for $p$
being expandable in a positive series in $z$.
We therefore try to fit it into a structure obtained from
$T_i$'s acting on configuration
$(-\mathsf{a},1,\mathsf{a},1)$:
\begin{equation} \begin{split}
{\bar \alpha}_1 &=-\mathsf{a}+2 n_1-2 n_2,\;
{\bar \alpha}_2=  1 +2 n_2-2 n_3,\\
 {\bar \alpha}_3&= \mathsf{a}+2 n_3-2 n_4,\quad {
 \bar \alpha}_4= 1 +2 n_4-2 n_1,
\label{ullate-ex}
\end{split}
\end{equation}
For ${\bar \alpha}_1= \frac{13}{2},  {\bar \alpha}_2= -1, {\bar \alpha}_3= -\frac{5}{2},  {
	\bar \alpha}_4= -1$, it is now easy to find a class of solutions
\[  n_2 = -1+n_3,\; n_1 = 1+n_4,\; \mathsf{a} = -\frac{5}{2}-2n_3+2n_4
\]
with $n_3,n_4$ being arbitrary integers. If we set f.i. $n_3=n_4=1$,
then $n_2=0$ and $\mathsf{a}=-5/2$ form the solution.
\end{exmp}
Note that relations \eqref{ullate-ex} are equivalent 
with 
\begin{equation}
{\bar \alpha}= \frac12  \left(\mathsf{a}/2 + n_3- n_4\right)^2, \;\;
{\bar \beta}=- \frac12 \left(-\mathsf{a}/2 + n_1- n_2 \right)^2,  \;\;
{\bar \gamma}= \epsilon \, (n_2-n_3-n_4+n_1),
\label{baralphas1aa}
\end{equation}
Setting $a=- \left(\mathsf{a}/2 + n_3- n_4\right)=
\mathsf{a}/2-n_1+n_2$ we can rewrite the above
as
\[
{\bar \alpha}= \frac12  \left(a +m\right)^2, \;\;
{\bar \beta}=- \frac12 \left(a \right)^2,  \;\;
{\bar \gamma}= \epsilon \, k\, ,
\]
with $m=n_1-n_2+n_3-n_4, k=n_2-n_3-n_4+n_1$ and
$k+m=2 n_1-2n_4$ being an even number (see also \cite{kitaev} 
or (I) on page \pageref{page:items}). 
\begin{exmp}
In this example instead of connecting the solution \eqref{oursol2example} 
to the seed solution with $(-\alpha_1,1,\alpha_1,1)$ we will rather 
take the polynomial solution \eqref{s1oursol2}
with $\alpha_1=-\frac{13}{2},\alpha_2=\frac{11}{2},
\alpha_3=-\frac{5}{2} ,
\alpha_4=\frac{11}{2}$ obtained by acting with $s_1$ on solution \eqref{oursol2example}
from \cite{ullate} 
and show that it can be obtained from polynomial solution \eqref{solution2}
with 
\[ (\alpha_1, \alpha_2, \alpha_3, \alpha_4) =
(\mathsf{a},1-\mathsf{a},\mathsf{a},1-\mathsf{a}) \, ,\]
by successive operations of translations operations $T_i$ each acting
$n_i$ times.
Recalling the actions of $T_i$ \eqref{Tialpha} we obtain the following
$4$ conditions for the solution \eqref{s1oursol2} to be obtained from 
the solution \eqref{solution2} by $T_i$'s each acting
$n_i$ times:
\[ \begin{split}
\mathsf{a}+2 n_1-2 n_2&= -\frac{13}{2}, \quad
(1-\mathsf{a})+2 n_2-2 n_3 = \frac{11}{2}\\
\mathsf{a}+2 n_3-2 n_4&= -\frac{5}{2} , \quad
(1-\mathsf{a})+2 n_4-2 n_1=  \frac{11}{2}\, ,
\end{split}\]
with a general  solution given in terms of arbitrary $n_3,n_4$:
\[ n_1 = n_3-1,\; n_2 = n_4+1,\; \mathsf{a} = -\frac{5}{2}+2 n_4-2n_3,
\]
that involves action by the shift operators equal to
\[
T_1^{-1+n_3}T_2^{1+n_4}T_3^{n_3}T_4^{n_4}= 
\left(T_1T_3\right)^{-1+n_3} \left(T_2T_4\right)^{1+n_4} 
T_3 T_4^{-1}\, .
\]
The above expression shows that there is no ambiguity 
related to the choice of $n_3$ and $n_4$ as  $(T_1T_3)^{-1+n_3}$ and 
$(T_2T_4)^{1+n_4}$  do not change the form of the solution.
Therefore for simplicity we eliminate the first two factors of the above 
expression by choosing :
\[ \; n_3-1=n_4+1=0\;
\;\to \;\; n_1=n_2=0,\; \;\mathsf{a}=-\frac{13}{2}, \]
and thus the action of shift operators \eqref{T1234} 
becomes that of $T_3 T_4^{-1}$. The action of the inverse operator
$T_4^{-1}=\pi^{-1} s_1s_2s_3$ on $p=q=z/2,
(\mathsf{a},1-\mathsf{a},\mathsf{a},1-\mathsf{a})$ is well defined and yields
\[ 
q=\frac12 z \frac{(z^2-4 \mathsf{a})}{(z^2-4 \mathsf{a}-4)}, \;
p=\frac12 z \frac{(z^2-4 \mathsf{a}+4)}{(z^2-4 \mathsf{a})} , \;
(\mathsf{a},1-\mathsf{a},2+\mathsf{a},-1-\mathsf{a})\, .
\]
Applying $T_3$ on the above expressions we get :
\[ \begin{split}
q&=
\frac{z(16\mathsf{a}^2+32\mathsf{a}-z^4)}{2(4\mathsf{a}+8+z^2)(-z^2+4\mathsf{a}+8)},\;
\;\;p=\frac{z (16\mathsf{a}^2+32\mathsf{a}-8 z^2-z^4)}{2(16\mathsf{a}^2+32\mathsf{a}-z^4)}\\
&(\mathsf{a},-1-\mathsf{a},4+\mathsf{a},-1-\mathsf{a})  \, , 
\end{split}\]
which for $\mathsf{a}=-\frac{13}{2}$ reproduces expression \eqref{s1oursol2}.
\end{exmp}

\subsubsection{Umemura polynomial solutions obtained from
$\pmb{j_i=\frac{z}{4} (1,1,1,1)}$ seed solution through action of the shift operators}

As follows from relations \eqref{N6arbitrary1recurj} applied to $N=4$
case we have the following recurrence relations 
\begin{equation}
\begin{split}
 j_{4 ,\, n+1} &=  T_3 ( j_{4 ,\, n})=  j_{3 \, n}- \frac{\mathsf{a}+2n}{ j_{3 ,\, n}+ j_{4 ,\, n}} \\
 j_{1 ,\, n+1} &=  T_3 ( j_{1 ,\, n})= j_{4 ,\, n}+ \frac{\mathsf{a}+2n}{ j_{3 ,\, n}+ j_{4 ,\, n}}
 - (1+2n) \frac{j_{3,\, n}+ j_{4 ,\, n}}{(j_{3 ,\, n}+ j_{4 ,\, n})
 (j_{4 ,\, n}+ j_{1 ,\, n})+ \mathsf{a}+2n} \, , 
 \end{split}
 \label{T3recurj}
\end{equation}
for transformations induced by $T_3$.

Since $\sum_{i=1}^4 j_{4 ,\, n}=z$ and $\sum_{i=1}^4 j_{4 ,\, n+1}=z$ 
we find for $T_3 (j_{2 \, n}+j_{3 ,\, n}) =z - T_3 (j_{1 \, n}+j_{4 ,\, n})$
\[
\begin{split}
T_3 (j_{2 \, n}+j_{3 ,\, n})&= j_{1 \, n}+j_{2 ,\, n}
+ (1+2n) \frac{j_{3,\, n}+ j_{4 ,\, n}}{(j_{3 ,\, n}+ j_{4 ,\, n})
 (j_{4 ,\, n}+ j_{1 ,\, n})+ \mathsf{a}+2n}\\
&= j_{1 \, n}+j_{2 ,\, n}
+ (1+2n) \frac{z- (j_{1,\, n}+ j_{2 ,\, n})}{(z- j_{1 ,\, n}- j_{2 ,\, n})
 (z- (j_{2 ,\, n}+ j_{3 ,\, n}))+ \mathsf{a}+2n} \, , 
 \end{split}
\]
which can be rewritten as 
\begin{equation}
 p_{n+1} =q_n +(2n+1)\frac{\left(z-q_{n}\right)}{d_n} \, , 
\label{pnp1}
\end{equation}
where for $q=j_1+j_2, p=j_2+j_3$ we introduced the following notation
\begin{equation}
q_n (z;\mathsf{a}) =T_3^n(q_0), \quad p_n (z;\mathsf{a}) =T_3^n(p_0),
\quad q_0=p_0=\frac{z}{2}  \, , 
\label{qndef}
\end{equation}
and 
\begin{equation}
d_{n} (z;\mathsf{a})  =(z-q_n)(z-p_n)+2 n+\mathsf{a} \, .
\label{dnza}
\end{equation}
Similarly from equations \eqref{N6arbitrary1recurj} we find
\[
T_3 (j_{1 \, n}+j_{2 ,\, n})= j_{1 \, n}+j_{4 ,\, n}
+ \frac{(\mathsf{a}+2n) }{(z- j_{1 ,\, n}- j_{2 ,\, n})}
- \frac{(\mathsf{a}+2n+1) }{z- T_3(z- j_{2 ,\, n}- j_{3 ,\, n})} \, , 
\]
that can be rewritten as
\begin{equation}
q_{n+1}=z-p_n+\frac{\mathsf{a}+2n}{z-q_n}
-\frac{\mathsf{a}+2n+1 }{p_{n+1}} = \frac{d_n}{z-q_n} -\frac{\mathsf{a}+2n+1 }{p_{n+1}} \, , 
\label{qnp1}
\end{equation}
and together with equation \eqref{pnp1} form two recurrence relations
for the canonical quantities $q_n,p_n$.
One finds from relations  \eqref{pnp1} and  \eqref{qnp1} that 
\[
\frac{q_n}{z-q_n} d_n = q_{n+1} p_{n+1} +\mathsf{a} \, ,
\] 
which shows that the quantity $d_n$ is useful in describing transition
from $p_n,q_n$ to $p_{n+1},q_{n+1}$. Indeed we will be able below to 
formulate the recurrence relation for Umemura polynomials based on
existence of alternative expressions \eqref{dniu} for $d_n$.

It is convenient to introduce the polynomials $U_n (z;\mathsf{a})$  to which we will 
refer as Umemura polynomials \cite{noumi98u,kajiwara} defined for
$n=0,1,2,3$ by 
\begin{align}
U_0 (z;\mathsf{a})&=1, \quad U_1 (z;\mathsf{a})=1\, ,\label{Uone} \\	
U_2 (z; \mathsf{a})&= z^2+4 \mathsf{a} \, , \label{Utwo}\\
U_3 (z;\mathsf{a} )&=z^6+12 z^4 \mathsf{a}+12 z^4+48 z^2 \mathsf{a}^2+96 z^2 \mathsf{a}
+192 \mathsf{a}^2+128 \mathsf{a}+64 \mathsf{a}^3 \, .
\label{Uthree}
\end{align}
Note that $U_n (z;\mathsf{a})= z^{n(n-1)}+ \ldots$ is a polynomial of the 
$n(n-1)$-th order.
In terms of the above polynomials we can express $q_1,p_1, d_1$
in the following way
\begin{equation}
	\begin{split}
q_1 (z;\mathsf{a}) &= \frac{z}{2} \frac{U_2 (z;\mathsf{a}) U_1 (z;\mathsf{a}+3)}{U_2(z;\mathsf{a}+1) U_1 (z;\mathsf{a}+2)}
, \quad \; p_1 (z;\mathsf{a})  = \frac{z}{2} \frac{U_2(z;\mathsf{a}+1)
U_1 (z;\mathsf{a})}{U_2(z;\mathsf{a}) U_1 (z;a +1)} \, ,\\
d_1 (z;\mathsf{a}) &= 
\frac{z^{2}}{4} \frac{U_{2} (z;\mathsf{a}+2) \,U_{2} (z;\mathsf{a}-1)}{U_{2} (z;\mathsf{a}+1)\,
U_{2} (z;\mathsf{a})}
+2  +\mathsf{a} =
\frac{1}{4}  \frac{U_3 (z;\mathsf{a}) U_1 (z;\mathsf{a}+1)}{U_2
(z;\mathsf{a}) U_2 (z;a+1)} \,.
	\end{split}
\label{pqone}
\end{equation}
The repeating action of $T_3$ operator on expressions \eqref{pqone} gives rise
to  :
\begin{equation}
	\begin{split}
q_n (z; \mathsf{a}) &= \frac{z}{2} \frac{U_n (z;\mathsf{a}+3) \,U_{n+1} (z;\mathsf{a})}{U_n
(z;\mathsf{a}+2)\, U_{n+1} (z;\mathsf{a}+1)},\; 		\quad
p_n (z;\mathsf{a}) = \frac{z}{2} \frac{U_n (z;\mathsf{a})\, U_{n+1}(z;\mathsf{a}+1)}{U_n
(z;\mathsf{a}+1)\, U_{n+1} (z;\mathsf{a})}\\
d_n (z;\mathsf{a}) &=
\frac{z^{2}}{4} \frac{U_{n+1} (z;\mathsf{a}+2) \,U_{n+1} (z;\mathsf{a}-1)}{U_{n+1} (z;\mathsf{a}+1)\, U_{n+1} (z;\mathsf{a})}
+2 n +\mathsf{a} \, .
\end{split}
	\label{pqT3n}
\end{equation}
Using the recurrence relations \eqref{pnp1},  \eqref{qnp1} one 
can alternatively express the quantity  $d_n =(z-q_n)(z-p_n)+2 n+\mathsf{a} $ as
\begin{equation}
	\begin{split}
d_{n} (z;\mathsf{a}) &= 
\frac14 \frac{U_{n} (z;\mathsf{a}+1) U_{n+2} (z;\mathsf{a}) }{U_{n+1} (z;\mathsf{a}) U_{n+1} (z;\mathsf{a}+1)}
\\&=\frac14 \frac{U_{n} (z;\mathsf{a}+2) U_{n+2} (z;\mathsf{a}-1)
}{U_{n+1} (z;\mathsf{a}) U_{n+1} (z;\mathsf{a}+1)}+(2n+1) \,. 
\end{split}	
\label{dniu}
\end{equation}
Comparing the bottom of expressions \eqref{pqT3n} with the two
expressions in equation \eqref{dniu} we obtain two alternative 
recurrence relations
for the Umemura polynomials which independently can be used to
generate higher level Umemura polynomials.

It is convenient at this point  to 
introduce the variable $x=\frac{z^{2}}{4}$
and polynomials 
\begin{equation}
W_n (x;\mathsf{a})= 2^{-n(n-1)} U_n (z;\mathsf{a}) \, ,
\label{un2wn}
\end{equation}
which satisfy two recurrence relations that follow from comparing expressions
\eqref{pqT3n} with \eqref{dniu} :
\begin{align}
W_{n-1} (x;\mathsf{a}+1) W_{n+1} (x;\mathsf{a})&= x W_{n} (x;\mathsf{a}+2)  W_{n}
(x;\mathsf{a}-1)\nonumber \\&
+ (2 n-2 +\mathsf{a})  W_{n} (x;\mathsf{a}) W_{n} (x;\mathsf{a}+1) 	
\label{recurWx}	 \\
W_{n-1} (x;\mathsf{a}+3) W_{n+1} (x;\mathsf{a})&= x W_{n} (x;\mathsf{a}+3)  W_{n} (x;\mathsf{a}) 
+\mathsf{a}  W_{n} (x;\mathsf{a}+2) W_{n} (x;\mathsf{a}+1) \, .
	\label{recurWxa}	 
\end{align}
Such redefined Umemura polynomials $W_n(x;\mathsf{a})$ are
given for $n=0,1,2,3, 5$ by 
\begin{align}
W_0 (x;\mathsf{a})&=1, \quad W_1 (x;\mathsf{a})=1 \, ,\label{Wone}\\	
W_2 (x; \mathsf{a})&= x+ \mathsf{a} \, ,\label{Wtwo}\\
W_3 (x;\mathsf{a} )&=x^3+3 x^2 \mathsf{a}+3 x^2+3 x \mathsf{a}^2+6 x \mathsf{a}
+ 3 \mathsf{a}^2+2 \mathsf{a}+ \mathsf{a}^3 
\label{Wthree}\\
&=(x+\mathsf{a})^{3} +3 (x+\mathsf{a})^{2} +2 \mathsf{a} \, ,\nonumber\\
W_4 (x;\mathsf{a}) &=
48\,\mathsf{a}+60\, x^3+x^6+12\, x^5+45\, x^4+144\, x\,\mathsf{a}+240\, x^2\,\mathsf{a}+300\, x\,\mathsf{a}^2+
6\, x^5\,\mathsf{a} +15\, x^4\,\mathsf{a}^2 \nonumber\\&+60\, x^4\,\mathsf{a}
+20\, x^3\,\mathsf{a}^3+120\, x^3\,\mathsf{a}^2+190\, x^3\,\mathsf{a}+15\, x^2\,\mathsf{a}^4+120\, x^2\,\mathsf{a}^3
+300\, x^2\,\mathsf{a}^2+60\, x\,\mathsf{a}^4\nonumber\\&+210\, x\,\mathsf{a}^3+6\, x\,\mathsf{a}^5+124\,\mathsf{a}^2
+120\,\mathsf{a}^3+12\,\mathsf{a}^5+\mathsf{a}^6+55\,\mathsf{a}^4\, ,
\label{Wfour}
\end{align}
from which higher polynomials can be obtained using recurrence
relations \eqref{recurWx} or \eqref{recurWxa}.
In addition, the polynomials $W_n (x;\mathsf{a})$ satisfy the 
identity 
\begin{equation}
 2 W_{n+1} (x, \mathsf{a} )   W_n (x;\mathsf{a}+1) -W_{n+1} (x, \mathsf{a}+1 ) W_n (x; \mathsf{a}) =
 W_{n+1} (x;\mathsf{a}-1) W_n (x;\mathsf{a}+2)\, ,
 \label{id1Wmemura}	
 \end{equation}
 established on basis of consistency
of the shift operator approach with various operators $T_i$ connected
via $\pi$. Although we have chosen arbitrarily to generate the recurrence relations
by acting with $T_3$ we could taken any other shift operator as a
starting point and be able to transfer from one formalism to another
by applying the automorphism $\pi$ through relation $ \pi T_i= T_{i+1} \pi$.
The identity \eqref{id1Wmemura} ensures that acting with any of the
shift operators $T_i, i=1,2,3,4$ on expressions \eqref{pqone} will give 
rise to  solutions that are still expressible in terms of Umemura
polynomials $U_n (z; \mathsf{a})$.
For example the repeating action of $T_1$ operator on 
expressions \eqref{pqone} yields :
\begin{subequations}
	\begin{align}
q_n^{(1)} (z;\mathsf{a}) &= \frac{z}{2} \frac{U_n(z,\mathsf{a}+1)
U_{n+1}(z,\mathsf{a}+2)}{U_n(z,\mathsf{a}+2) U_{n+1}(z,\mathsf{a}+1)}, \label{T1nqn} \\
		p_n^{(1)} (z;\mathsf{a})  &= \frac{z}{2} \frac{U_n(z,\mathsf{a}+2) 
		U_{n+1}(z,\mathsf{a}-1)}{U_n(z,\mathsf{a}+1) U_{n+1}(z,\mathsf{a})}.
	\end{align}
\end{subequations}

Consider again equation  \eqref{pqT3n} for $q_n(z;\mathsf{a})$
and plug $q_n$ into expression $y= (q/z) (q/z-1)^{-1}$ for solution of
Painlev\'e V equation derived in
subsection \ref{subsection:H2pain}. 
After some simple algebra we  find :
\[y = \frac{W_n (x;\mathsf{a}+3) \,W_{n+1} (x;\mathsf{a} )}{W_n (x;\mathsf{a}+3) \,W_{n+1} (x;\mathsf{a}
)-2 W_n (x;\mathsf{a}+2)\, W_{n+1} (x;\mathsf{a}+1)}\, .
\]
Using the identity \eqref{id1Wmemura} to rewrite the denominator we obtain
\begin{equation}
y = - \frac{W_n (x;\mathsf{a} +3) \,W_{n+1} (x;\mathsf{a} )}{W_n
(x;\mathsf{a}+1) \,W_{n+1} (x;\mathsf{a}+2 )} \, ,
\label{yUmemura}
\end{equation}
for $(\mathsf{a},1-\mathsf{a}- 2n, \mathsf{a}+2 n,1-\mathsf{a})$ with the Painlev\'e parameters:
\[
{\bar \alpha}= \frac12 (\frac{\mathsf{a}}{2}+n)^2,\;\; {\bar \beta}=- 
\frac12 (\frac{\mathsf{a}}{2})^2,\;\;
{\bar \gamma}= - \frac{1}{\epsilon } n\, ,
\]	
agreeing with the solution (Ib) given at the beginning of section \ref{section:rational}.

Consider now solution \eqref{T1nqn} 
generated by acting $n$ times with 
the shift operator $T_1$. The parameters $\alpha_i$ for this solution
are equal to
$( \mathsf{a} + 2n , 1-\mathsf{a}, \mathsf{a}, 1-\mathsf{a}-2 n)$.
Plugging the above $q(z)$ into  expression $y= (q/z) (q/z-1)^{-1}$
and using the identity  \eqref{id1Wmemura} we get
\begin{equation}
y  
=- \frac{W_n (x;\mathsf{a} +1) \,W_{n+1} (x;\mathsf{a}+2 )}{W_n (x;\mathsf{a}+3) \,W_{n+1} (x;\mathsf{a}
)}\, ,
\label{y1Umemura}
\end{equation}
with the
Painlev\'e V parameters 
\[
{\bar \alpha}=  \frac12 (\frac{\mathsf{a}}{2})^2,\;\; 
{\bar \beta}=- \frac12 (\frac{\mathsf{a}}{2}+n)^2,\;\;
{\bar \gamma}= n, \; {\bar \delta}= - \frac12\, ,
\]	
that agree with the solution (Ia) given at the beginning of section 
\ref{section:rational}
for  the Painlev\'e V variable $t = 2 x $. 

The fact that the above $y$ satisfies the Painlev\'e V equation
is equivalent to the Umemura polynomials $W_n (x,\mathsf{a}) $ satisfying the 
$\sigma$-type of relation,
which can be given a form of a Toda like equation:
\[
\frac{W_{n-1} (x;\mathsf{a}) W_{n+1} (x;\mathsf{a}) }{ W_n^2 (x;\mathsf{a})}=
  x +\mathsf{a} +3 (n-1)+ 2 \frac{d}{d x } x \frac{d}{d x } \ln W_n
  (x;\mathsf{a}) \, .
\]
Next we define quantity:
\begin{equation}
\omega_a=\frac{W_n (x;\mathsf{a}) \,W_{n+1} (x;\mathsf{a}+1 )}{W_n
(x;\mathsf{a}+1) \,W_{n+1} (x;\mathsf{a} )} \,  - 1,
\label{omegaa}	
\end{equation}
where we suppressed dependence on $n$ on the left hand side.
It is interesting to notice that as follows from applications of all 
three identities \eqref{recurWx}, \eqref{recurWxa} and \eqref{id1Wmemura} 
$\omega_a$ satisfies a discrete Painlev\'e II equation \cite{kajiwara}:
 \begin{equation}
\omega_{a-1}+\omega_{a+1}=
\frac{2}{x} \frac{1}{1-\omega_a^2} \left( n + (1-n-a) \omega_a\right)\, .
\label{dpII}
\end{equation}
See \cite{aratyn93} for an  early observation that B\"acklund
transformations of continuous models can give rise to a discrete
structure.

\subsection{Action of the shift operators on $\pmb{j_i=\frac{z}{2}
(1,1,-1,1)}$ solution in  \eqref{solution1}}
\label{subsection:item3}

By acting with $T_1^{n_1} T_2^{n_2} T_3^{n_3} T_4^{n_4}$   
on  $j_i= ({z}/{2}) (1,1,-1,1)$ from equation \eqref{solution1}
with 
$\alpha_i=(\mathsf{a},0,0,2-\mathsf{a})$  
we will arrive, in principle, at the following  
parameters of the final configuration
\[ \begin{split}
{\alpha}_1 &= \mathsf{a} +2 n_1-2 n_2,\;{\alpha}_2= 2 n_2-2 n_3,\\
 {\alpha}_3&= 2 n_3-2 n_4,\quad { \alpha}_4= 2-\mathsf{a} +2
n_4-2 n_1\, ,
\end{split}
\]
or 
\begin{equation}
{ \alpha}_1 = \mathsf{a} +2 n_1-2 n_2,\; 
{ \alpha}_3= 2 n_3-2 n_4,\; 
\frac{{ \alpha}_2-{ \alpha}_4}{2}=\frac{\mathsf{a}}{2} -1 +n_2-n_3-n_4+
n_1, \;  
n_i \in  \mathbb{Z}\, .
\label{baralphas2}
\end{equation}
However not all of the shift transformations are well defined when acting on 
  $j_i\;=\; ({z}/{2}) (1,1,-1,1)$. 
Since $j_2+j_3=0$ and $j_3+j_4=0$ we see from the definition
\eqref{Tinewj} that actions of $s_2,s_3$ involve divisions by zero
and therefore are not allowed.  
{}Recalling the definitions \eqref{trans} and \eqref{inverseT} we accordingly need to
exclude $T_2,T_3$ and $T_3^{-1} , T_4^{-1}$ as these operators contain
$s_3$ and $s_2$ transformations at the positions to the right.
Because the shift
operators in \eqref{trans} and \eqref{inverseT} 
contain ordered products of neighboring B\"acklund transformations
of the type $s_{i+1} s_i$ the divergence is only generated by the
$s_i$ located to the right. If the result of acting by $s_i$ is not
divergent then acting  with $s_{i+1}$ would not be divergent as follows
from the definition \eqref{Tinewj}.

Accordingly, to avoid divergencies we will only consider  the operators 
$T_1^{n_1} T_4^{n_4} T_2^{-n_2}$ with $n_2,n_4 \in \mathbb{Z}_{+}$ and
$n_1 \in \mathbb{Z}$.

Indeed 
one can verify that $T_2^{-1}=s_2s_3 \pi^{-1} s_1$ is permissible and
generates 
\begin{equation}
\begin{split}
T_2^{-1}&: q=z, p=0 \to q=z, p= \frac{2z}{\mathsf{a} -z^2}, \, 
(2+\mathsf{a},-2,0,2-\mathsf{a})\\
T_2^{-n}&: q=z, p=0 \to q_n=z, p_n=\frac{2n z R_{n-1} (\mathsf{a};z)}{R_{n}
(\mathsf{a};z) },  \, 
(2n+\mathsf{a},-2n,0,2-\mathsf{a})\, ,
\label{T2n}
\end{split}\end{equation}
where $R_{n} (\mathsf{a};z) $ is found to satisfy the recurrence
relation:
\[R_{n+1} (\mathsf{a};z)= 2 n z^2 R_{n-1}
(\mathsf{a};z)+(-z^2+2n+\mathsf{a}) R_{n} (\mathsf{a};z), \quad
n=1,2,{\ldots} \, ,\]
with $R_0 (\mathsf{a};z)=1$. The solution to this recurrence relation is
given by
\begin{equation}
R_{n}
(\alpha_1;z)= \sum_{r=0}^{n} \, \binom{n}{r} \, (\mathsf{a})_{r,2}\,
(-z^2)^{n-r}, \quad n=0,1,2, {\ldots} \, ,
\label{pochhammer}
\end{equation}
where we used the Pochhammer k-symbol $(x)_{n,k}$ defined as 
$(x)_{n,k}=x(x+2)(x+2k) \cdots (x+(n-1)k)$. 
We notice that $R_{n} (\mathsf{a};z)$ can be expressed as a function of
$x=-z^2/2$ and in terms of $x$ it holds that $d R_{n} (\mathsf{a};x)/d x
= 2n R_{n-1} (\mathsf{a};x)$. Thus we find that $p_n$ from equation \eqref{T2n}
satisfies $p_n/z=f_2/z= d  (\ln R_{n} (\mathsf{a};x))/d x$.
Based on discussion around equation \eqref{Umn} from subsection
\ref{subsection:riccati} we expect that $R_{n} (\mathsf{a};x)$ is related 
to Kummer's polynomial $U(-n, 1-n-\mathsf{a}/2, x)$.
Indeed an explicit calculation of expression \eqref{pochhammer}  yields 
$R_{n} (\mathsf{a};x)= 2^n x^{n+\mathsf{a}/2}
\, U (\frac{\mathsf{a}}{2}, \frac{\mathsf{a}}{2}+n+1, -\frac{z^2}{2})$,
which according to relation \eqref{Umn} is equal (up to an overall
constant) to $U(-n, 1-n-\mathsf{a}/2, x)$, a solution to the
Kummer's equation \eqref{kummereq}  with $a= \alpha_2/2=-n$, 
$b=(\alpha_2+\alpha_4)/2= -n+1-\mathsf{a}/2$. Here we obtained
this solution through acting $n$-th times with  $T_2^{-1}$ on the first-order
solution \eqref{solution1}. Since the Kummer's functions found many applications 
in e.g. solvable quantum mechanics, atomic physics and critical phenomena among other fields the fact 
that as shown above their form  can be reproduced by action of the shift operators should be of potential interest for these applications and efforts to expand them.

The shift operator $T_1$   essentially acts  as an identity
\[
T_1 : q=z,\, p=0,\, \alpha_i=(\mathsf{a},0,0,2-\mathsf{a})\; \longrightarrow\;
q=z, \, p=0,\, \alpha_i=(2+\mathsf{a},0,0,-\mathsf{a})\, ,
\]
its only action is to increase $\mathsf{a} \to \mathsf{a}+2$.

Let us now take a closer look at the action of $T_4$ on $q=z, p=0$. Acting once with $T_4$ yields :
\begin{equation}
q_1= z- \frac{2 z}{z^2-\mathsf{a}+2}=T_4 (q_0),\;\; p_1=0, \;\;\, (\mathsf{a},0,-2,4-\mathsf{a})\, ,
\label{T4onq0}
\end{equation}
Acting $n$ times with $T_4$ on $q_0=z, p=0$ we get $q_n= T_4^n (q_0)$
that satisfies the recurrence relation
\begin{equation}
q_n=  z - \frac{2 n z }{z q_{n-1}+2 n -\mathsf{a}},\quad\;
 (\mathsf{a},0,-2n,2(n+1)-\mathsf{a})\, ,
\label{recurrenceT4}
\end{equation}
the corresponding expression for $p_n$ is
\[ p_n = q_{n-1}+\frac{2n-\mathsf{a}}{z}-\frac{2n}{z-q_n}=0\, ,
\]
where the zero on the right hand side follows from the recurrence
relation \eqref{recurrenceT4} connecting $q_n, q_{n-1}$. 

It we assume that $F_{n-1}= q_{n-1}/z$ satisfies the Riccati equation
\eqref{ric1} for $i=1$ and $\alpha_3=-2(n-1)$ then it follows that 
$F_{n}= q_{n}/z$ with $q_n$ determined through the recurrence relation
\eqref{recurrenceT4} will satisfy the same Riccati equation
\eqref{ric1} for $\alpha_3=-2n$. Since for  $q_0=z$ the function
$F_0=q_0/z=1$ satisfies the Riccati equation
\eqref{ric1} for $\alpha_3=0$ this concludes the induction proof
for $q_n$ being equal to $z F_{\mathsf{a},
\alpha_3=-2n}$ where $ F_{\mathsf{a},\alpha_3}$ is given by
expression \eqref{fwhittaker} in terms of Whittaker functions.

Based on the above discussion we can rewrite equation \eqref{baralphas2}
as
\begin{equation}
	{ \alpha}_1 = \mathsf{a} +2 n_1+2 n_2,\; 
	{ \alpha}_3= -2 n_4,\; 
	\frac{{ \alpha}_2-{ \alpha}_4}{2}=\frac{\mathsf{a}}{2} -1 -n_2-n_4+
	n_1, \;  
n_2,n_4 \in  \mathbb{Z}_{+}, n_1 \in \mathbb{Z}\, ,
	\label{baralphas2b}
\end{equation}
after making a transformation $n_2 \to -n_2$.

Accordingly, equation \eqref{baralphas2b} gives rise to
\begin{equation}
	{\bar \alpha}= \frac12  \left( - n_4\right)^2, \;\;
	{\bar \beta}=- \frac12 \left(\mathsf{a}/2 + n_1+n_2 \right)^2,  \;\;
	{\bar \gamma}= \epsilon (\frac{\mathsf{a}}{2} -1 -n_4-n_2+
	n_1), \;  
	\label{baralphas3a}
\end{equation}
or after elimination of an  arbitrary constant $\mathsf{a}$ :
\begin{equation}
	{\bar \alpha}= \frac12  \left( - n_4\right)^2, \;\;
{\bar \beta}=- \frac12 \left( \epsilon {\bar \gamma}+1+n_4+2n_2 \right)^2,
	\label{baralphas3b}
\end{equation}

After learning how the solution \eqref{solution1} transforms under a product of fundamental shift
operators we turn our attention to action of these operators on solutions that can be obtained
from \eqref{solution1} by an automorphism $\pi$. Acting with $\pi$ and $\pi^2$ on \eqref{solution1}
we get respectively, $j_i= ({z}/{2}) (1,1,1,-1)$ with 
$(2-\mathsf{a},\mathsf{a},0,0)$ and $j_i= ({z}/{2}) (-1,1,1,1)$ with
$(0,2-\mathsf{a},\mathsf{a},0)$ as seeds configurations.

For $j_i= ({z}/{2}) (1,1,1,-1)$ we see that $j_3+j_4=0$ and $j_4+j_1=0$.
Thus comparing with relations \eqref{Tinewj} we recognize that 
the B\"acklund transformations $s_3, s_3 \pi^{-1}, s_4, s_1 \pi$ would 
involve divisions by zero. Accordingly among the eight shift operators listed
in \eqref{trans} and \eqref{inverseT} we need to discard $T_4,T_3,
T_4^{-1},  T_1^{-1}$  that contain the above mentioned B\"acklund transformations 
in the positions to the right. Accordingly we will only act with
$T_1^{n_1} T_3^{-n_3} T_2^{n_2}$ with 
$n_1,n_3 \in  \mathbb{Z}_{+}, n_2 \in \mathbb{Z}$ generating the
following transformations of $(2-\mathsf{a},\mathsf{a},0,0)$ :
\begin{equation}
\begin{split}
{ \alpha}_1 &=2-\mathsf{a}+2 n_1-2 n_2,\;{ \alpha}_2=  
\mathsf{a} +2 n_2+2n_3,\\
 { \alpha}_3&= -2 n_3,\quad { \alpha}_4=  -2 n_1,
 \;\; n_1,  n_3 \in \mathbb{Z}_{+},\; n_2 \in \mathbb{Z}\, ,
 \label{pisolution1}
\end{split}
\end{equation}
The Painlev\'e parameters corresponding to \eqref{pisolution1} are:
\[
{\bar \alpha}= \frac12 (n_3)^2, \;\;
{\bar \beta}=- \frac12 \left(1-\mathsf{a}/2 + n_1- n_2 \right)^2,  \;\;
{\bar \gamma}=\epsilon (\frac{\mathsf{a}}{2}  +n_2+
n_1-n_3), \; 
\]
or \[
{\bar \alpha}= \frac12 (n_3)^2, \;\;
{\bar \beta}=- \frac12 \left( \epsilon {\bar \gamma} -1 -2 n_1+n_3 \right)^2,
\]
with $n_1,n_3$ being positive integers or zero. The above equation is 
similar to relation \eqref{baralphas3b}.

For $j_i= ({z}/{2}) (-1,1,1,1)$ we see that $j_1+j_4=0$ and $j_1+j_2=0$.
We conclude from relations \eqref{Tinewj} that 
the B\"acklund transformations $s_1, s_3 \pi^{-1}, s_4, s_1 \pi, s_2 \pi$ would 
involve divisions by zero. We therefore need to exclude $T_4,T_1,
T_2^{-1},  T_1^{-1}$ among the eight shift operators listed
in \eqref{trans} and \eqref{inverseT}.
The action with the remaining shift operators
$T_2^{n_2} T_4^{-n_4} T_3^{n_3}$ with 
$n_2,n_4 \in  \mathbb{Z}_{+}, n_3 \in \mathbb{Z}$ generates the
following transformation of $(0,2-\mathsf{a},\mathsf{a},0)$ :
\begin{equation} 
\begin{split}
{ \alpha}_1 &=-2 n_2,\;{ \alpha}_2= 2-\mathsf{a}+ 2 n_2-2 n_3,\\
 { \alpha}_3&= \mathsf{a} +2 n_3+2 n_4,\quad { \alpha}_4= -2 n_4,
\;\;  n_2, n_4 \in \mathbb{Z}_{+}, \,n_3 \in \mathbb{Z}\, ,
\label{pi2solution1a}
\end{split}
\end{equation}
The Painlev\'e parameters corresponding to \eqref{pi2solution1a} are:
\[
{\bar \alpha}= \frac12  \left( \frac{\mathsf{a}}{2}+n_3+n_4 \right)^2, \;\;
{\bar \beta}=- \frac12 \left( - n_2 \right)^2,  \;\;
{\bar \gamma}= \epsilon \,(1-\frac{\mathsf{a}}{2}  +n_2-n_3+n_4), \; 
\]
or
\begin{equation}
{\bar \alpha}= \frac12  \left( \epsilon \,{\bar \gamma}-1-n_2-2 n_4 \right)^2, \;\;
{\bar \beta}=- \frac12 \left( n_2 \right)^2,  \;\;
\label{baralphas4b}
\end{equation}
with $n_2,n_4$ being positive integers or zero.
Relations \eqref{baralphas3b} and \eqref{baralphas4b} constitute 
item (III) on page \pageref{page:items1}.

\begin{exmp} Let us now consider the following example with solution
taken from \cite{ullate} :
\begin{equation}
q= \frac{z(z^4-14z^2+63)}{z^4-18z^2+99},
\;\; p= \frac{z^6-21 z^4 +189 z^2-693}{z(z^4-14 z^2+63)},\;
\alpha_1=7,\alpha_2=6, \alpha_3=-4
\label{oursol3}
\end{equation}
Expression for $p$ has a pole which can be removed by applying $s_1$. 
Applying $s_1$ we get
\begin{equation}
q = \frac{z(z^4-14z^2+63)}{z^4-18z^2+99},
\;\; p=z,
\;\;
\alpha_1=-7,\alpha_2=13, \alpha_3=-4, \, \alpha_4=0\, .
\label{oursol3a}
\end{equation}
We will match it with the initial configuration of \eqref{wata2} with
$p=z,q=z$ and $(2-\mathsf{a}, \mathsf{a},0,0)$
on which we can act with $T_2^{n_2}, T_3^{-n_3}, T_1^{n_1}$ (but not
$T_3^{+1}$)
to get :
\[ \begin{split}
{\alpha}_1 &=-7= 2-\mathsf{a} +2 n_1-2 n_2,\;{\alpha}_2= 13=
\mathsf{a} +2 n_2+2 n_3,\\
 {\alpha}_3&=-4=  -2 n_3,\quad {\alpha}_4=0=  -2 n_1, \; n_1, n_3 \in \mathbb{Z}_{+},\; n_2 \in \mathbb{Z}\, .
\end{split}
\]
We choose $\mathsf{a}=9, n_1=n_2=0, n_3=2$ to get the desired result.
One can show for the corresponding combination of shift operators that
$T_3^{-2}=\pi^2 s_1 s_2 s_3 s_4 s_1 s_2$ and acting with such operator
on $p=z,q=z$ and $\alpha_i=(-7, 9,0,0)$
one reproduces easily the solution  \eqref{oursol3a}. Alternatively, we
can obtain this solution as a special function solution when we
recognize that for the condition $\alpha_4=0$ from equation 
\eqref{oursol3a} the Hamilton equations \eqref{hameqs} are solved by 
$p=z$, which when inserted in the first equation in \eqref{hameqs} 
reduce this equation to the Riccati equation 
$ z q_z = -z q(q-z)+(1-\alpha_1-\alpha_3)  q +\alpha_1 z $ solved by
\[
q = \frac{\alpha_3}{z} \frac{{\rm  WhittakerM}
(\frac{\alpha_3}{4}-\frac{\alpha_1}{4}+1, -\frac12+\frac{\alpha_1}{4}+
\frac{\alpha_3}{4}, \frac{z^2}{2})}{
{\rm  WhittakerM}(\frac{\alpha_3}{4}-\frac{\alpha_1}{4},
-\frac12+\frac{\alpha_1}{4}+\frac{\alpha_3}{4}, \frac{z^2}{2})}
+\frac{z^2-\alpha_3}{z}\,.
\]
Inserting $\alpha_1=-7, \alpha_3=-4$ we recover from the above
expression the rational solution  \eqref{oursol3a}.
\end{exmp}

By comparing with results in \cite{kitaev} we conclude that 
acting with shift operators on the first-order polynomial solutions of
$N=4$ dressing chain produces all rational solutions of the associated
Painlev\'e system. We therefore conjecture that the same technique
will produce all rational solutions for higher even $N$ and discuss
realization of this statement for $N=6$ in the next section.

\section{Special function and rational solutions of $N=6$ equations}
\label{section:N=6}
\subsection{Reductions of $N=6$ Hamilton Equations \eqref{q1p2eqs}}

Recall that in subsection \ref{subsection:riccati} we considered 
$N=4$ solutions \eqref{wata3}
with $\alpha_i=(\mathsf{a},0,0,2-\mathsf{a})$. Having the parameters 
$\alpha_2$ or $\alpha_3$ set to zero resulted in 
 $N=4$ Hamilton equations \eqref{hameqs} being reduced to a single
 Riccati equation. For example for
 $\alpha_3=0$ the Hamilton equations \eqref{hameqs} are solved by
 $q=z$ and a solution of the Riccati equation $z p_z = z p (p-z) - (1-\alpha_1) p
 +\alpha_2 z$. Similarly for $\alpha_2=0$ the Hamilton equations 
 \eqref{hameqs} are solved by $p=0$ and a solution of the Riccati equation
$z q_z = z q(q-z)+(1-\alpha_1-\alpha_3)  q +\alpha_1 z$.
Accordingly we determined a class of special function solutions to the
Painlev\'e V equation that became rational solutions when
the $\alpha_i$ parameters coincided with orbits of 
$(\mathsf{a},0,0,2-\mathsf{a})$ obtained by an action of appropriate
shift operators.

In this subsection we will carry out a similar discussion for the $N=6$
case investigating conditions for presence of 
the special function solutions to
the  Hamilton equations \eqref{q1p2eqs}.
The Hamilton equations \eqref{q1p2eqs} represent four coupled nonlinear 
third-order differential equations. 
Setting  to zero various components of $\alpha_i$
introduces connections between $q_i,p_i, i=1,2$  and accordingly reduces
a number of coupled nonlinear equations. Imposing three
constraints on parameters of $N=6$ Hamilton system \eqref{q1p2eqs}
reduces the system to only one solvable second-order Riccati equation 
with a special function solution.
The three constraints emerge when the two of $j_i$ are
negative as in solutions \eqref{N6polynomialsols2}-\eqref{N6polynomialsols5}.

When the reduced systems are realized on orbits
of shift operators $T_i^{n_i}$ acting on seeds solutions 
\eqref{N6polynomialsols2}-\eqref{N6polynomialsols5} all these 
Riccati solutions become rational solutions
parameterized by integers $n_i$.

\subsubsection{One-constraint reductions of $N=6$ Hamilton Equations}

We will proceed by listing possible conditions on $\alpha_i$
parameters together with expressions for those $q_i,p_i, i=1,2$ that solve 
the reduced equations \eqref{q1p2eqs} obtained as a result of imposing 
constraints.
For examples the formula  :
\begin{equation}
\alpha_6=0 \quad \longrightarrow \quad p_2= z- p_1\, ,
\label{alpha6zero}
\end{equation}
means that inserting the condition $\alpha_6=0$ 
into the last two equations for $p_1,p_2 $
in \eqref{q1p2eqs} causes each of them to 
reduce to one identical equation for $p_1$ :
\[
z p_{1, \, z}= p_1 (z-p_1) (2 q_2 -2 q_1-z)+z \alpha_2 -
p_1 (1-\alpha_1-\alpha_3-\alpha_5)\, ,
\]
with $p_2=z- p_1$. The reduced system of the remaining 
three Hamilton equations only depends on three variables
$q_1,q_2,p_1$ after imposition of one single constraint.

We list below other  single constraints and the corresponding simple solutions 
for quantities entering equations \eqref{q1p2eqs} :
\begin{align}
\alpha_5&=0 \quad \longrightarrow \quad q_2= z\, ,
\label{alpha5zero} \\
\alpha_4&=0  \quad \longrightarrow \quad p_2= 0\, ,
\label{alpha4zero} \\
\alpha_3&=0  \quad \longrightarrow \quad q_1=q_2\, ,
\label{alpha3zero} \\
\alpha_2 &=0  \quad \longrightarrow \quad p_1=0\, ,
\label{alpha2zero} \\
\alpha_1 &=0  \quad \longrightarrow \quad q_1=0\, .
\label{alpha1zero}
\end{align}

\subsubsection{Multi-constraint reductions of $N=6$ Hamilton
Equations }

One can combine the above single constraints of $\alpha_i$ parameters into a set of two and more 
constraints. As we will see below the set of three constraints
leads to the constrained system described by a single Riccati equation.

Imposing two constraints leads as a rule to two coupled nonlinear equations
but not always  
equations that are quadratic in underlying variables.

Let us first consider the following example of two constraints:
\begin{equation}
\alpha_6=0 \;\&\;\; \alpha_5 =0 \quad \longrightarrow \quad p_1+p_2=z,
\;q_2= z\, ,
\label{alpha35zero}
\end{equation}
that combines $p_1+p_2=z$  that follows from $\alpha_6=0$ and
relation $q_2=z$  that follows from $\alpha_5=0$. 
Imposing these two relations we 
can rewrite the Hamiltonian equations only in terms of e.g.
$p_2, q_1$ entering cubic non-linear equations :
\begin{equation}\begin{split}
z p_{2, \, z}&= (z- p_2)p_2 (2 q_1-z)
+z \alpha_4 -p_2 (1-\alpha_1-\alpha_3)\, ,\\
z q_{1, \, z}&=  q_1(z-q_1) (z-2 p_2) +z \alpha_1 +q_1 (1-\alpha_1-\alpha_3)\, .
\end{split}
\label{q1p2red25}
\end{equation}

For the two constraints:
 \begin{equation}
 \alpha_4=0 \;\&\;\; \alpha_3+\alpha_5 =0 \quad \longrightarrow \quad
 p_2=0, \; q_1= z\, ,
 \label{alpha43+5zero}
 \end{equation}
the remaining variables $p_1,q_2$ enter two coupled second-order equations :
\begin{equation} \begin{split}
z p_{1\, z} &= -z p_1 (z-p_1) + z \alpha_2-
p_1 (1-\alpha_1)\, ,\\
z q_{2\, z}&=z( z-q_2) (2p_1-q_2)+z(\alpha_1+\alpha_3) +
q_2(1-\alpha_1)\, .
\end{split} 
\label{p1q2redeqs}
\end{equation}
Only the first equation is a Riccati equation solvable in terms of
Kummer/Whittaker functions.

Next consider the two constraints
\begin{equation}
 \alpha_6=0 \;\&\;\; \alpha_1+\alpha_5 =0 \quad \longrightarrow \quad
 p_2=z-p_1, \; q_2= q_1+z\, .
 \label{alpha61+5zero}
 \end{equation}
The two remaining equations for $q_1,p_1$ are found to be 
\begin{equation}
\begin{split}
z q_{1, \, z}&= z q_1 (2 p_1-z) -z q_1^2
+z \alpha_1 +q_1 (1-\alpha_3)\, ,\\
z p_{1, \, z}&= z p_1 (z-p_1)
+z \alpha_2 -p_1 (1-\alpha_3)\, .
\end{split}
\label{q1p2redeqs}
\end{equation}
The second equation among equations \eqref{q1p2redeqs} is a regular
Riccati equation but the first one is a coupled Riccati equation.
We will see below in Example \ref{example-2.6orbits} that the 
coupled equations \eqref{p1q2redeqs} and
\eqref {q1p2redeqs} become fully solvable on orbits of the shift
operators.

Combining together  conditions into three conditions
yields one single second-order 
Riccati equation emerging from such reduction process.
\begin{equation}
\alpha_2=0 \;\&\; \alpha_5 =0 \;\&\; \alpha_6=0 \quad \longrightarrow \quad
q_2= z, \, p_1=0,p_2=z\, .
\label{alpha256zero}
\end{equation}
In this case there only remains one  Riccati
equation for the remaining variable $q_1$ :
\begin{equation} z q_{1\, z} =-zq_1(z-q_1)+z  \alpha_1
+q_1 (1-\alpha_1-\alpha_3)\, .
\label{T4q1}
\end{equation}

Replacing $\alpha_2$ with $\alpha_4$ in \eqref{alpha256zero} yields 
\begin{equation}
\alpha_2=0 \;\&\; \alpha_5 =0 \;\&\; \alpha_6=0 \quad \longrightarrow \quad
q_2= z, \, p_1=z,p_2=0\, ,
\label{alpha456zero}
\end{equation}
with a Riccati
equation for $q_1$ 
\begin{equation} z q_{1\, z} =zq_1(z-q_1)+z  \alpha_1
+q_1 (1-\alpha_1-\alpha_3)\, .
\label{T3nq1}
\end{equation}

Similarly the three constraints
\begin{equation}
\alpha_6=0 \;\&\; \alpha_4 =0 \;\&\; \alpha_3+\alpha_5=0 \quad 
\longrightarrow \quad  p_1=z, \; p_2=0,\;  q_1=z\, ,
\label{alpha4635zero}
\end{equation}
leave only one Riccati equation for $q_2$ : $z q_{2, \, z}= z (z-q_2)^2 
+z^2 (z-q_2) +z (\alpha_1+\alpha_3) +q_2 (1-\alpha_1)$.
Plugging $f_3=q_2-z$ we get  a simple looking Riccati equation for $f_3$:
\begin{equation}
z f_{3\, z}= -z^2 f_3 +z f_3^2+z \alpha_3+f_3 (1-\alpha_1)\, .
\label{f3T2n}
\end{equation}
A similar case is that of three constraints with $\alpha_3$ replaced by
$\alpha_1$ : 
\begin{equation}
\alpha_6=0 \;\&\; \alpha_4 =0 \;\&\; \alpha_1+\alpha_5=0 \quad 
\longrightarrow \quad  p_1=z, \; p_2=0,\;  q_2=q_1+z\, ,
\label{alpha4615zero}
\end{equation}
which leaves only one Riccati equation for $q_1$ : $z q_{1\, z} =  z q_1 (z-q_1)  + z \alpha_1+
q_1(1-\alpha_3)$.

Further we also list the three constraints:
\begin{equation}
\alpha_6=0 \;\&\; \alpha_5 =0 \;\&\; \alpha_3=0 \quad 
\longrightarrow \quad  q_1=z, \, p_2+p_1=z,\; q_2=z\, ,
\label{alpha653zero}
\end{equation}
 As seen before $\alpha_6=0$ leads to  $p_2=z-p_1$ and $\alpha_5=0$
leads to $q_2=z$. One of the remaining Hamilton equations is 
$z q_{1, \, z}= q_1 (z-q_1)(2p_1-z) +z \alpha_1+q_1 (1-\alpha_1)$
with the solution $q_1=z$, which when inserted in equation 
for $p_1$ gives Riccati equation : $z p_{1, \, z}= -z p_1 (z-p_1) 
+z \alpha_2-p_1 (1-\alpha_1)$.

Another example of three independent constraints:
\begin{equation}
\alpha_6=0 \;\&\; \alpha_3 =0 \;\&\; \alpha_2=0 \quad 
\longrightarrow \quad  p_1=0, \, p_2=z\, .
\label{alpha632zero}
\end{equation}
The $N=6$ Hamilton equations \eqref{q1p2eqs} give then for the remaining
quantities $q_1,q_2$: 
\[
\begin{split}
z q_{1\, z} &= -z q_1 (q_2-q_1) +zq_1(z-q_2)+z  \alpha_1
+q_1 (1-\alpha_1-\alpha_5)\, , \\
z q_{2\, z}& = z (z-q_2) (q_2-q_1) +zq_1(z-q_2)
+z \alpha_1+q_2 (1-\alpha_1-\alpha_5) \, .
\end{split}
\]
Taking the difference of the above two equations yields equation for
$q_2-q_1$ which is solved for $q_2=q_1$. Thus we are left with
one Riccati equation for $q_1$ : $z q_{1\, z} = zq_1(z-q_1)+z  \alpha_1
+q_1 (1-\alpha_1-\alpha_5)$.

Another case of three constraints
\begin{equation}
\alpha_5=0 \;\&\; \alpha_4 =0 \;\&\; \alpha_3=0 \quad 
\longrightarrow \quad  q_1=z, \, p_2=0,\; q_2=z\, ,
\label{alpha453zero}
\end{equation}
lead to one single Riccati equation for the remaining quantity $p_1$ :
\begin{equation}
z p_{1, \, z}= -z p_1 (z-p_1)
+z \alpha_2 -p_1 (1-\alpha_1) \,.
\label{2-7eq-p1}
\end{equation}

As seen above the three constraints reduce the four Hamiltonian equations in \eqref{q1p2eqs}
to one Riccati equation for the remaining variable. 
As expected imposing all four constraints 
applied on the four Hamiltonian equations in \eqref{q1p2eqs}
leads only to trivial solutions: 
\begin{equation}
\alpha_6=0 \;\&\;\alpha_5=0 \;\&\; \alpha_4 =0 \;\&\; \alpha_3=0 \quad 
\longrightarrow \quad  p_1=z, \, p_2=0, \,q_1= q_2=z\, .
\label{alpha4536zero}
\end{equation}

As we will see below in example \ref{example-2.6orbits} 
there are cases of two constraints with two remaining
Riccati equations that decouple under special circumstances when the
parameters are chosen to coincide with the orbits of the shift
operators.

\subsection{The orbit construction of rational solutions for \pmb{$N=6$}}
\label{subsection:N6case}
In this section we apply the technique introduced in previous
sections to the case of $N=6$ for which we already found the 
first-order polynomial solutions in equations 
\eqref{N6polynomialsols1}-\eqref{N6polynomialsols5}.

As found in subsection \ref{subsection:kova} for the $N=6$ case 
after the appropriate actions by $s_1$ 
and $s_3$ the variables $p_i, i=1,2$ can be expanded in 
positive power series that do not contain pole singularities. Such 
rational solutions 
can then be reproduced by actions of the shift operators on 
solutions \eqref{N6polynomialsols1}-\eqref{N6polynomialsols5} or
 \eqref{N6polynomialsols1pq}-\eqref{N6polynomialsols5pq}.

\subsubsection{Umemura polynomial solutions for \pmb{$N=6$}}

In this subsection we will apply the fundamental shift operator 
techniques to 
\begin{enumerate}
\item[I)] The seed solution \eqref{N6polynomialsols1} with all
components $j_i=z/N=z/6$
\item[II)] The seed solution \eqref{N6polynomialsols2} with one of the 
components being negative and equal to  $-z/(N-2)=-z/4$
\end{enumerate}
The case I) will require new class of Umemura polynomials \eqref{gukn}
with the leading order term being $z^{n(n-p)}, p=1, n/2-1,n/2$ with
the last two cases being new. In case II) we will be able to essentially reduce the problem to that of $N=4$ and
express the solutions in terms of regular Umemura polynomials
with the leading order term being $z^{n(n-1)}$.

Case I). Recall the relevant $N=6$ shift operators from definitions
\eqref{N6shifts} and \eqref{inverseT6}.
For solution \eqref{N6polynomialsols1} 
with all $j_i=z/6, i=1,2,3,4,5,6$ 
it holds that  $j_i+j_{i+1} \ne 0$
for all $i=1,2,3,4,5,6$. Thus all $s_i$ transformations acting via
relation \eqref{Tinewj} are well defined and
action by 
\[ T_1^{n_1}  T_2^{n_2}  T_3^{n_3} T_4^{n_4}  T_5^{n_5}  T_6^{n_6} 
, \qquad n_i \in \mathbb{Z}, i=1,2,3,4,5,6\, ,
\]
produces rational solutions with the transformed ${\bar \alpha}_i$:
\[
(\mathsf{a}+2n_1 -2n_2,\frac23-\mathsf{a}+2n_2 -2n_3, \mathsf{a}
+2n_3-2n_4,\frac23-\mathsf{a}+2n_4-2n_5,
\mathsf{a}+2n_5-2n_6,\frac23-\mathsf{a}+2n_6-2n_1)\, .
\]
We can rewrite the above action of the shift operators as follows
\[ \begin{split}
 T_1^{n_1}  T_2^{n_2}  T_3^{n_3} T_4^{n_4}  T_5^{n_5}  T_6^{n_6} &=
(T_1T_3T_5)^{n_1} (T_2 T_4  T_6)^{n_2} T_3^{n_3-n_1} T_5^{n_5-n_1}
T_4^{n_4-n_2}T_6^{n_6-n_2}\\
&=(T_1T_3T_5)^{n_1} (T_2 T_4  T_6)^{n_2} (T_3T_4)^{k_{+}} (T_3T_4^{-1})^{k_{-}} 
(T_5T_6)^{m_{+}}(T_5T^{-1}_6)^{m_{-}}\, ,
\end{split}\]
where
\[ \begin{split}
k_{+}&= \frac12 ( n_3-n_1+n_4-n_2), \qquad k_{-} = \frac12 (
n_3-n_1-n_4+n_2)\,,\\
m_{+}&= \frac12 ( n_5-n_1+n_6-n_2), \qquad m_{-}
= \frac12 ( n_5-n_1-n_6+n_2) \, .
\end{split}\]
One can easily prove that $(T_1T_3T_5)^{n}$ only shifts the parameter
$\mathsf{a}$: $\mathsf{a} \to \mathsf{a}-n $ without changing the
functional form of the solution \eqref{N6polynomialsols1pq}.
Similarly $(T_2T_4T_6)^{n}$ only shifts the parameter
$\mathsf{a}$: $\mathsf{a} \to \mathsf{a}+n $ leaving the solutions 
\eqref{N6polynomialsols1pq} unchanged.

For  $(T_3T_4)^{k_{+}}$ we find that it results in
$(\alpha_2-\alpha_4)/2=-2 k_{+}, (\alpha_3-\alpha_5)/2=0$, while 
for $(T_3T_4^{-1})^{k_{-}} $ we obtain 
$(\alpha_2-\alpha_4)/2=0, (\alpha_3-\alpha_5)/2=2 k_{-}$.

For  $(T_5T_6)^{m_{+}}$ we find that it results in
$(\alpha_6-\alpha_4)/2=2 m_{+}, (\alpha_3-\alpha_5)/2=0$, while 
for $(T_5T_6^{-1})^{m_{-}} $ we obtain 
$(\alpha_6-\alpha_4)/2=0, (\alpha_3-\alpha_5)/2=-2 m_{-}$.

For $N=6$ we introduce the following 
notation:
\begin{equation}
U_{k,n} (z;\mathsf{a} ) =  z^{(n+k)(n)} + \ldots, \quad k>0, \, n>0 \, ,
\label{gukn}
\end{equation}
which generalizes Umemura polynomials of the type $U_{1,n-1} (z;a )=
z^{n(n-1)}+ {\ldots} $
seen in the previous section for $N=4$.
These new  Umemura polynomials take the following special values for
$n=0,1,2 $:
\begin{align}
U_{-1,1} (z;\mathsf{a} ) &= 1 \,, \;U_{-1,0} (z;\mathsf{a} ) =U_{0,0} (z;\mathsf{a} )=1 \,, \nonumber\\
U_{1,-1}(z;\mathsf{a} )&= 1, \quad U_{1,0}(z;\mathsf{a} )= 1  \,,\\
U_{1,1} (z;\mathsf{a}  )& = z^2+9 \mathsf{a} \label{N6T1two}  \,,\\
U_{0,2} (z;\mathsf{a}  ) &=  U_{3,1} (z;\mathsf{a})
 =  z^4+18\,z^2\,\mathsf{a}+24\,z^2+162\,\mathsf{a}+81\,\mathsf{a}^2+72\label{N6T1threetwo}\\
U_{1,2} (z;\mathsf{a}  ) &= 
648+2106\, a+540\,z^2\,\mathsf{a}+2187\,\mathsf{a}^2+252\,z^2+33\,z^4 \nonumber\\&
+z^6+27\,z^4\,\mathsf{a}+243\,z^2\,\mathsf{a}^2+729\,\mathsf{a}^3 	
	\label{N6T1three}
\end{align}
and enter the following expressions for solutions we obtained by acting
once with the shift operator $T_1$  on the $n=0$ configuration \eqref{N6polynomialsols1} with 
$q_1=p_1=p_2=z/3$ and $q_2=2z/3$ and 
$\alpha_i=(\mathsf{a},2/3-\mathsf{a},\mathsf{a},2/3-\mathsf{a},\mathsf{a},2/3-\mathsf{a})$
:
\begin{align}
q_1^{(1)}&=\frac{z}{3} \frac{U_{-1,1} (z; \mathsf{a}) \,U_{0,2} (z; \mathsf{a}+2/3) }
{U_{-1,1} (z; \mathsf{a}+2/3) \,U_{0,2} (z; \mathsf{a})   }= \,\frac13\,
	\frac {z \left( {z}^{4}+18\,{z}^{2}\mathsf{a}+36\,{z}^{2}+270\,\mathsf{a}+81\,{\mathsf{a}}^{2}+216 \right) }
	{{z}^{4}+18\,{z}^{2}\mathsf{a}+24\,{z}^{2}+162\,\mathsf{a}+81\,{\mathsf{a}}^{2}+72} 
\label{q1N6n1}\\
p_1^{(1)} &=\frac{z}{3} \frac{U_{-1,1} (z; \mathsf{a}+2/3) \,U_{1,1} (z; \mathsf{a}-2/3)}
{U_{-1,1} (z; \mathsf{a}) \,U_{1,1} (z; \mathsf{a})} 
=\,\frac13\, \frac {z \left( {z}^{2}+9\,\mathsf{a}-6 \right) }{{z}^{2}+9\,\mathsf{a}}
\label{p1N6n1}\\	
q_2^{(1)}  &=
\,\frac{2z}{3}\, \frac{	U_{1,2} (z; \mathsf{a})}{U_{0,2} (z; \mathsf{a})\,U_{1,1} (z;
\mathsf{a}+2/3)}	\nonumber\\
	&=\,\frac23\, \frac {z \left( 648+2106\,\mathsf{a}+540\,{z}^{2}a+2187\,{a}^{2}+252\,{z}^{2}+33\,{z}^{4}+{z}^{6}+27\,{z}^{4}\mathsf{a}+243\,{z}^{2}{\mathsf{a}}^{2}+729\,{\mathsf{a}}^{3}			 \right) }{ \left( {z}^{2}+9\,\mathsf{a}+6 \right)  \left( {z}^{4}+18\,{z}^{2}\mathsf{a}+24\,{z}^{2}+162\,\mathsf{a}+81\,{\mathsf{a}}^{2}+72 \right) } \nonumber\\
\label{q2N6n1}\\
p_2^{(1)} &=
	\frac{z}{3} \frac{U_{1,1} (z; \mathsf{a}+2/3) \,U_{0,2} (z; \mathsf{a}-4/3)\,  }
	{U_{1,1} (z; \mathsf{a}) \,U_{0,2} (z; \mathsf{a}-2/3)\,}  =
\,\frac13\,\frac {z  
	(z^2+9 \mathsf{a}+6) (z^4+18 z^2 \mathsf{a}-54\mathsf{a}+81\mathsf{a}^2)}
	{ \left( {z}^{4}+18\,{z}^{2}\mathsf{a}+12\,{z}^{2}+54\,\mathsf{a}+81\,{\mathsf{a}}^{2} \right) 
			\left( {z}^{2}+9\,\mathsf{a} \right) } \nonumber\\	
	\label{p2N6n1}
\end{align}
The repeated action $n$-th times with the shift operator $T_1$ on 
\eqref{N6polynomialsols1} with 
$q_1=p_1=p_2=z/3$ and $q_2=2z/3$ can be described by generalization
of \eqref{q1N6n1}-\eqref{p2N6n1} given by:
\begin{align}
q_1^{(n)}&=\frac{z}{3} \frac{U_{n-2,n} (z; \mathsf{a} )\, U_{n-1,n+1} (z; \mathsf{a} +2/3) }
{U_{n-2,n} (z; \mathsf{a} +2/3) \,U_{n-1,n+1} (z; \mathsf{a} )   }   \label{q1N6n}\\
p_1^{(n)} &=\frac{z}{3} \frac{U_{n-2,n} (z; \mathsf{a} +2/3) \,U_{n,n} (z; \mathsf{a} -2/3)}
{U_{n-2,n} (z; \mathsf{a} ) \,U_{n,n} (z; \mathsf{a} )} \label{p1N6n}\\	
q_2^{(n)}  &=
\,\frac{2z}{3}\, \frac{	U_{1,2n} (z; \mathsf{a} )}{U_{n-1,n+1} (z; \mathsf{a} )\,U_{n,n} (z;
\mathsf{a} +2/3)}	
\label{q2N6n}\\
p_2^{(n)} &=
	\frac{z}{3} \frac{U_{n,n} (z; \mathsf{a} +2/3) \,U_{n-1,n+1} (z; \mathsf{a}-4/3)\,  }
	{U_{n,n} (z; \mathsf{a}) \,U_{n-1,n+1} (z; \mathsf{a}-2/3)\,}\, , \label{p2N6n}
\end{align}
where $U_{k,n} (z;\mathsf{a} )$ polynomials of the type shown in
equation \eqref{gukn}.

Case II). For the solution \eqref{N6polynomialsols2} with $j_i=\frac{z}{4} (1,1,1,1,1,-1)$,
it holds that  $j_5+j_6=0, j_6+j_1=0$ and
that makes $s_i, s_{i-1} \pi^{-1}, s_{i+1} \pi$ with $i=5,6$ 
ill-defined. Accordingly $T_5, T_6, T_1^{-1}, T_6^{-1}$ are ill-defined.
Rational solutions will be produced from the seed solution
\eqref{N6polynomialsols2} by action of
\[
T_1^{n_1}  T_2^{n_2} T_3^{n_3} T_4^{n_4} T_5^{-n_5} , \;\;
n_1, n_5\in \mathbb{Z}_{+} ,\; n_2,n_3, n_4 \in \mathbb{Z}\, , 
\]
that yields the orbit parameters:
\begin{equation}
(\mathsf{a}+2n_1-2n_2,1-\mathsf{a}+2n_2-2n_3, \mathsf{a}+2n_3-2n_4
,1-\mathsf{a}+2n_4+2n_5, -2 n_5,-2 n_1)\, .
\label{2-5orbits}
\end{equation}
When we set $n_5=0$ in expression \eqref{2-5orbits} we obtain the
condition \eqref{alpha5zero} with $\alpha_5=0, q_2=z$. Inserting these
conditions into $N=6$ Hamilton equations \eqref{q1p2eqs} 
we find that $q_1,p_1$ satisfy separately 
the $N=4$ Hamilton equations \eqref{hameqs} although  they still
couple to $p_2$ but only in  the  last of equations \eqref{q1p2eqs}.
Explicitly, we find the solutions in terms of Umemura polynomials from
subsection \ref{subsection:item1&2}:
\begin{equation}
\begin{split}
q_{1}^{(n)} (z; \mathsf{a}) &=T_1^n (q_{1}^{(0)})= 
\frac{z}{2} \frac{U_n (z;-\mathsf{a}+2) \,
U_{n+1} (z;-\mathsf{a}+3)}{U_n
(z;-\mathsf{a}+3)\, U_{n+1} (z;-\mathsf{a}+2)},\\
p_{1}^{(n)} (z;\mathsf{a}) &=  T_1^n (p_{1}^{(0)})= 
\frac{z}{2} \frac{U_n (z;-\mathsf{a}+3)\, 
U_{n+1}(z;-\mathsf{a})}{U_n
(z;-\mathsf{a}+2)\, U_{n+1} (z;-\mathsf{a}+1)} \, ,
\end{split}
\label{p1nq1n}
  \end{equation}
with $q_{1}^{(0)} = p_{1}^{(0)} =z/2$. 
Given these two solutions one finds the expression for $p_{2}^{(n)}$ by
solving the corresponding equation of $p_2$ among $N=6$ Hamilton 
equations \eqref{q1p2eqs}. Since $p_{2}^{(n)}$  can be obtained by repeated
actions of the shift operator it  is given by a ratio of polynomials
as illustrated in the Example given below.

\begin{exmp}
Let us consider an orbit generated by $T_2^n$.  Plugging $n_2=n$ and
$n_1=n_3=n_4=n_5=0$ into  expression \eqref{2-5orbits}
we find that $\alpha_6=\alpha_5=0$ as in \eqref{alpha35zero}. 
The expressions for $p_2$ found by applying $T_2^n$ on 
solutions \eqref{N6polynomialsols2} with $p_2=z/2, q_1=z/2$
are :
\[ \begin{split}
p_2 (n=1, \mathsf{a}, z)&= \frac{z}{2} \frac{4+z^2 -4 \mathsf{a}}
{z^2 -4 \mathsf{a}+8} , \;\;
p_2 (n=-1,\mathsf{a}, z)= \frac{z}{2} \frac{z^2 -4 \mathsf{a}+4}
{z^2 -4 \mathsf{a}} \, , \\
q_1 (n=1, \mathsf{a}, z)&= \frac{z}{2} \frac{z^2 -4 \mathsf{a}+8}
{z^2 -4 \mathsf{a}+4} , \;\;
q_1 (n=-1, \mathsf{a}, z)= \frac{z}{2} \frac{-8+ z^2 -4 \mathsf{a}}
{-4+z^2 -4 \mathsf{a}}\, , 
\end{split}
\]
and one can  verify that they satisfy the 
relevant reduced Hamilton equations \eqref{q1p2red25} for $p_2, q_1$ with
$\alpha_1= \mathsf{a}- 2n, \alpha_2=1-\mathsf{a}+2n,
\alpha_3=\mathsf{a},
\alpha_4=1-\mathsf{a}$. 
\end{exmp}

\subsubsection{Riccati solutions for $N=6$}
In this subsubsection we will consider solutions constructed out of the
seed solutions with two negative $j_i$ components.
First consider solution given in \eqref{N6polynomialsols3} with 
$j_i=\frac{z}{2} (1,1,1,1,-1,-1)$ and $j_4+j_5=0, j_6+j_1=0$. 
These conditions render $s_i, s_{i-1} \pi^{-1}, s_{i+1} \pi, i=4,6$ 
ill-defined. Using these arguments we find that 
$T_4, T_6, T_1^{-1}, T_5^{-1}$ are ill-defined. Also 
by inspection we find that $T_3, T_5, T_2^{-1}, T_6^{-1}$ are ill-defined
as well.
Rational solutions will be produced from the seed solution
\eqref{N6polynomialsols3} by action of
\[
 T_1^{n_1} T_2^{n_2} T_3^{-n_3} T_4^{-n_4},\; \;
n_1,n_2 , n_3, n_4 \in \mathbb{Z}_{+} , 
\]
that yields
\begin{equation} (\mathsf{a}_3 +2n_1-2 n_2 ,2-\mathsf{a}_3-2n_2+2n_3, 
\mathsf{a}_3-2 n_3+n_4,-2 n_4,-\mathsf{a}_3,-2n_1)\, .
\label{2-6orbits}
\end{equation}
\begin{exmp}
Consider an orbit generated by $T_2^n$ obtained by inserting $n_2=n$ 
and $n_1=n_3= n_4=0$ into the above expression \eqref{2-6orbits}. 
This results in $\alpha_4=0, \alpha_6=0, \alpha_3+\alpha_5=0$,
which are the three  constraints shown in \eqref{alpha4635zero}.
The corresponding Riccati equation \eqref{f3T2n} becomes :
\[
z q_{2\, z}=-3 z^2 q_2 +z q_2^2 +2 z^3 +z(2 \mathsf{a}_3-2 n) +
q_2(1-\mathsf{a}_3+2n)\, ,
\] 
after inserting $\alpha_1=-2n +\mathsf{a}_3$.
Solving this equation for $q_2$ we get:
\[
q_2= z+  \frac{\mathsf{a}_3 }{z} +\frac{2}{z} \frac{
{\rm WhittakerW}(\frac{n}{2}+\frac{\mathsf{a}_3}{4}+1, 
-\frac12-\frac{n}{2}+\frac{\mathsf{a}_3}{4},
\frac{z^2}{2})}{
 {\rm WhittakerW}(\frac{n}{2}+\frac{\mathsf{a}_3}{4}, -\frac12-\frac{n}{2}
+\frac{\mathsf{a}_3}{4},
\frac{z^2}{2})} \, .
\]
For $n=0$ we obtain  $q_2=2z$, and next
\[ 
n=1, \quad q_2= 2 z\frac{(-\mathsf{a}_3+1+z^2)}{
(-\mathsf{a}_3+2+z^2)}\, , \]
\[n=2, \quad q_2=2z \frac{(-4 \mathsf{a}_3+4+\mathsf{a}_3^2-2 \mathsf{a}_3 z^2+2 z^2+z^4)}{
(-6 \mathsf{a}_3+8+\mathsf{a}_3^2-2 \mathsf{a}_3 z^2+4 z^2+z^4)} \, .\]
This is in agreement with results obtained by acting explicitly by 
$T_2^n, n=0,1,2$ on solution \eqref{N6polynomialsols3pq}

Similar considerations are involved in a study of an orbit generated
by $T_3^{-n}$ obtained by inserting $n_3=n$ 
and $n_1=n_2= n_4=0$ into expression \eqref{2-6orbits}. 
This results in $\alpha_4=0, \alpha_6=0, \alpha_1+\alpha_5=0$
which are the three  constraints shown in \eqref{alpha4615zero}.
Plugging $\alpha_1=\mathsf{a}_3$ and or $\alpha_3= \mathsf{a}_3-2n$
into the Riccati equation for $q_1$ shown below equation \eqref{alpha4615zero}
we find  a solution :
\begin{equation}
q_1(n,\mathsf{a}_3,z) = -2n \frac{{\rm WhittakerM}(-\frac{\mathsf{a}_3}{4}-\frac{n}{2}+1, -\frac12+\frac{\mathsf{a}_3}{4}
-\frac{n}{2}, \frac{z^2}{2})}{z {\rm WhittakerM}(-\frac{\mathsf{a}_3}{4}
-\frac{n}{2}, -\frac12+\frac{\mathsf{a}_3}{4}-\frac{n}{2}, \frac{z^2}{2})}
+\frac{(z^2+2n)}{z} \, , \label{2-6q1nT3}
\end{equation}
with $q_1 (0,\mathsf{a}_3,z) =z$ and
\[  q_1 (1,\mathsf{a}_3,z) =
z\frac{\mathsf{a}_3+z^2}{\mathsf{a}_3-2+z^2}, \;
q_1 (2,\mathsf{a}_3,z) =z \frac{-2\mathsf{a}_3+\mathsf{a}_3^2+2\mathsf{a}_3 z^2+z^4}{\mathsf{a}_3^2-
6\mathsf{a}_3+8+2\mathsf{a}_3 z^2-4 z^2+z^4} \,.
\]
\end{exmp}

\begin{exmp}
\label{example-2.6orbits}
The two examples we will here consider involve systems that are
characterized by two conditions imposed on the parameters $\alpha_i$.
Such situation leads to a system of reduced Hamilton equations
quadratic in canonical variables.
In examples shown here the reduced 
Hamilton equations consist one simple Riccati equation and one quadratic 
equation with coupled underlying canonical variables.
However when  $\alpha_i$ parameters are those
of an orbit 
\eqref{2-6orbits} the coupled Hamilton equations system  separates
into two independent and solvable  Riccati equations.

First, we consider an $T_1^n$ orbit which is obtained by inserting $n_1=n$ 
and $n_2=n_3= n_4=0$ into the above expression \eqref{2-6orbits}. The
orbit configuration agrees with  the two constraints of 
\eqref{alpha43+5zero} and with the corresponding coupled Hamilton equations 
\eqref{p1q2redeqs} of which only the first equation is a Riccati
equation, which after inserting  
$\alpha_1=\mathsf{a}_3+2n$ yields
\[
z p_{1\, z} = -z p_1 (z-p_1) + z (2-\mathsf{a}_3)-
p_1 (1-\mathsf{a}_3-2n)\, ,
\]
with solution : 
\begin{equation} 
p_1(n,\mathsf{a}_3,z) = \frac{n \mathsf{a}_3}{z} \frac{{\rm WhittakerW}(-\frac12+\frac{n}{2}-\frac{\mathsf{a}_3}{4},
\frac{\mathsf{a}_3}{4}+\frac{n}{2}, \frac{z^2}{2})}{{\rm WhittakerW}(\frac12+\frac{n}{2}-\frac{\mathsf{a}_3}{4},
\frac{\mathsf{a}_3}{4}+\frac{n}{2}, \frac{z^2}{2})}+\frac{(z^2-2 n)}{z}\, ,
\label{p1na3T1}
\end{equation}
for which we find for $n=0,1,2$:
\begin{equation}
\begin{split}
p_1 (n=0,\mathsf{a}_3,z) &= z, \;\quad p_1 (n=1,\mathsf{a}_3,z)  =  z
\frac{\mathsf{a}_3+z^2-2}{\mathsf{a}_3+z^2}\\
p_1 (n=2,\mathsf{a}_3,z) &=
z \frac{-2\mathsf{a}_3+\mathsf{a}_3^2+2 z^2\mathsf{a}_3+z^4-4
z^2}{2\mathsf{a}_3+\mathsf{a}_3^2+2 z^2\mathsf{a}_3+z^4}\, .
\label{p1T1}
\end{split}
\end{equation}
However for $\alpha_1=\mathsf{a}_3+2n$ it appears that effectively 
equations \eqref{p1q2redeqs} decouple.
We can namely define $q_1$ such that 
\[ q_1(n,\mathsf{a}_3,z)=q_2(n,\mathsf{a}_3,z)-z\, ,
\]
that satisfies the Riccati equation :
\[
z q_{1 \, z}= -z q_1 (z-q_1)+z (-\mathsf{a}_3)+q_1 (1+\mathsf{a}_3+2n)\, ,
\]
with solution:
\[
q_1= \frac{2}{z}\frac{ {\rm WhittakerW}(\frac{n}{2}-\frac{\mathsf{a}_3}{4}+1, 
\frac12+\frac{\mathsf{a}_3}{4}+\frac{n}{2}, \frac{z^2}{2})}
{{\rm WhittakerW}(\frac{n}{2}-\frac{\mathsf{a}_3}{4}, 
\frac12+\frac{\mathsf{a}_3}{4}+\frac{n}{2}, \frac{z^2}{2})}-
\frac{\mathsf{a}_3}{z}\, ,
\]
which gives
explicitly the values
\[
\begin{split}
q_1(n=0,\mathsf{a}_3,z) &=z,\;\quad
q_1 (n=1,\mathsf{a}_3,z)= z
\frac{\mathsf{a}_3+z^2}{\mathsf{a}_3+z^2+2}\\
q_1(n=2,\mathsf{a}_3,z)&=z
\frac{2\mathsf{a}_3+\mathsf{a}_3^2+2z^2\mathsf{a}_3+z^4}{
6\mathsf{a}_3+8+\mathsf{a}_3^2+2z^2\mathsf{a}_3+4 z^2+z^4}\, ,
\end{split}
\]
that reproduces $q_2 (n,\mathsf{a}_3,z)$ after adding $z$.

Quite similar behavior will take place for an orbit  $T_4^{-n}$ 
obtained by inserting $n_4=n$ 
and $n_1=n_2=n_3=0$ into the above expression \eqref{2-6orbits}.
Here $\alpha_i$ parameters satisfy two conditions :
$\alpha_6=0$ and $\alpha_1+ \alpha_5=0$ which coincide with expression
\eqref{alpha61+5zero}.
The two $N=6$
Hamilton equations \eqref{q1p2eqs} for remaining variables
 $q_1,p_1$ shown in \eqref{q1p2redeqs}   are such that 
the first equation contains a coupling between these two variables.
although the second equation is a regular
Riccati equation.
We consider the case of $\alpha_3= \mathsf{a}_3+2n$ and
$\alpha_2=2-\mathsf{a}_3, \alpha_1=\mathsf{a}_3$.
The solution to the second equation in \eqref{q1p2redeqs} is :
\begin{equation} p_1(n,\mathsf{a}_3,z) = -\frac{n \mathsf{a}_3}{z} 
\frac{{\rm WhittakerW}(-\frac12+\frac{n}{2}-\frac{\mathsf{a}_3}{4},
\frac{\mathsf{a}_3}{4}+\frac{n}{2}, -\frac{z^2}{2})}{
{\rm WhittakerW}(\frac12+\frac{n}{2}-\frac{\mathsf{a}_3}{4},
\frac{\mathsf{a}_3}{4}+\frac{n}{2}, -\frac{z^2}{2})}+\frac{(z^2+2 n)}{z}\, ,
\label{p1na3T4}
\end{equation}
Plugging $n=0,1,2$ into \eqref{p1na3T4} we get:
\begin{equation}
\begin{split}
p_1 (n=0,\mathsf{a}_3,z) &= z, \; p_1 (n=1,\mathsf{a}_3,z)  =  z
\frac{-\mathsf{a}_3+z^2+2}{-\mathsf{a}_3+z^2}\\
p_1 (n=2,\mathsf{a}_3,z) &=
z \frac{-2\mathsf{a}_3+\mathsf{a}_3^2-2 z^2\mathsf{a}_3+z^4-4
z^2}{2\mathsf{a}_3+\mathsf{a}_3^2-2 z^2\mathsf{a}_3+z^4}\, ,
\label{p1T4}
\end{split}
\end{equation}
It further holds for the particular values
 $\alpha_3= \mathsf{a}_3+2n$ and
$\alpha_2=2-\mathsf{a}_3, \alpha_1=\mathsf{a}_3$ that characterize 
the orbit that $q_1$ from equation \eqref{q1p2eqs} 
solves the Riccati equation
\begin{equation}
z q_{1\, z} = z q_1(z-q_1) - z \mathsf{a}_3+
q_1(1+\mathsf{a}_3+2n) \, ,
\label{2-6T4q1eq}
\end{equation}
and solution is  
\[
q_1(n,\mathsf{a}_3,z) = -2n \frac{{\rm WhittakerM}(\frac{\mathsf{a}_3}{4}-\frac{n}{2}+1, -\frac12-
\frac{\mathsf{a}_3}{4}
-\frac{n}{2}, \frac{z^2}{2})}{z {\rm WhittakerM}(\frac{\mathsf{a}_3}{4}
-\frac{n}{2}, -\frac12-\frac{\mathsf{a}_3}{4}-\frac{n}{2}, \frac{z^2}{2})}
+\frac{(z^2+2n)}{z} \, .
\]
For $n=0,1,2$ the above $q_1(n,\mathsf{a}_3,z)$ is equal to 
equal to 
\[\begin{split}
q_1 (n=0,\mathsf{a}_3,z)&=z,\quad \; q_1 (n=1,\mathsf{a}_3,z) =
z\frac{-\mathsf{a}_3+z^2}{-\mathsf{a}_3-2+z^2}, \\
q_1 (n=2,\mathsf{a}_3,z)&=z \frac{2\mathsf{a}_3+\mathsf{a}_3^2-2\mathsf{a}_3
z^2+z^4}{\mathsf{a}_3^2+
6\mathsf{a}_3+8-2\mathsf{a}_3 z^2-4 z^2+z^4}\, ,
\end{split}\]
which agrees with separate calculation involving the relevant shift
operator.
\end{exmp}

Next, consider solution given in \eqref{N6polynomialsols4}
$j_i=\frac{z}{2} (1,1,1,-1,1,-1)$ with 
the corresponding parameters $\alpha_i=(2-\mathsf{a}_2,\mathsf{a}_2, 0
,0,0,0)$ 
for which $j_3+j_4=0, \, j_4+j_5=0, \,j_5+j_6=0, j_6+j_1=0$.
With these quantities being zero we are not permitted to act with
$s_i, s_{i-1} \pi^{-1}, s_{i+1} \pi$ with $i=3,4,5,6$ on 
$j_i$ in \eqref{N6polynomialsols4} in order to avoid division by zero.
For these reasons we can not act with the shift operators
$T_i, T_{i+1}^{-1}, i=3,4,5,6 $ on solution given in \eqref{N6polynomialsols4}.
We can therefore only act with 
\[
 T_1^{n_1} T_2^{n_2}  T_3^{-n_3},\; n_1, n_3 \in \mathbb{Z}_+, \, n_2 \in \mathbb{Z}\, ,
 \]
that yields
\begin{equation}
(2-\mathsf{a}_2+2n_1-2n_2,\mathsf{a}_2-n_2+2n_3, 0-2 n_3
,0,0,0)\, .
\label{2-7orbits}
\end{equation}
\begin{exmp}
The action with the shift operator $T_1^{n}$ is implemented
by  setting $n_1=n, n_2=n_3=0$. Then the parameters $\alpha_i$
automatically satisfy the three conditions $\alpha_5=0, \alpha_3=0, \alpha_4=0$
as in equation \eqref{alpha453zero}.
The single Riccati equation for the remaining quantity $p_1$ is
given in equation \eqref{2-7eq-p1}.
Inserting  $\alpha_1=2 n+2-\mathsf{a}_2$ into equation \eqref{2-7eq-p1}
leads to rational solution given by :
\[
p_1(n,\mathsf{a}_2,z)= \frac{2}{z} \frac{{\rm WhittakerW}(\frac{\mathsf{a}_2}{4}
+\frac{n}{2}+1, \frac12+\frac{n}{2}-\frac{\mathsf{a}_2}{4}, \frac{z^2}{2})
}{{\rm WhittakerW}(\frac{\mathsf{a}_2}{4}+\frac{n}{2}, \frac12 +\frac{n}{2}
-\frac{\mathsf{a}_2}{4}, \frac{z^2}{2})}
+\frac{\mathsf{a}_2}{z}\, ,
\]
for which we find for $n=0,1,2$:
\begin{equation}
\begin{split}
p_1 (n=0,\mathsf{a}_2,z) &= z, \;\quad p_1 (n=1,\mathsf{a}_2,z)  =  z
\frac{-\mathsf{a}_2+z^2}{-\mathsf{a}_2+2+z^2}\\
p_1 (n=2,\mathsf{a}_2,z) &=
z \frac{-2\mathsf{a}_2+\mathsf{a}_2^2-2 z^2\mathsf{a}_2+z^4}
{8-6\mathsf{a}_2+\mathsf{a}_2^2+4 z^2- 2 z^2\mathsf{a}_2+z^4} \,.
\label{2-7-p1T1}
\end{split}\end{equation}
They agree with expressions obtained directly by acting with $T_1$ on 
solution given in \eqref{N6polynomialsols4}.

\end{exmp}

Finally, we consider solution given in \eqref{N6polynomialsols5}
\[j_i=\frac{z}{2} (1,1,-1,1,1,-1)
,\; \quad \alpha_i=(2-\mathsf{a}_4, 0,0,\mathsf{a}_4,0,0)\, ,
\]
for which $j_2+j_3=0, \, j_3+j_4=0, \,j_5+j_6=0, j_6+j_1=0$.
Accordingly  $s_i, s_{i-1} \pi^{-1}, s_{i+1} \pi$ with $i=2,3,5,6$ 
will involve division with zero.
This observation excludes $T_2, T_3, T_5, T_6, T_1^{-1},$ $T_3^{-1},
 T_4^{-1} , T_6^{-1} $. Thus we generate rational solutions by acting 
 with 
 \[
 T_1^{n_1} T_4^{n_4}  T_2^{-n_2} T_5^{-n_5}, n_1, n_2,n_4,n_5  \in \mathbb{Z}_+
 \, ,\]
that produces the parameter change
\begin{equation}
(2-\mathsf{a}_4+2n_1+2n_2, -2n_2,-2n_4,\mathsf{a}_4+2n_4+2n_5,-2n_5,-2n_1)\,.
\label{2-8orbits}
\end{equation}
\begin{exmp}
Here we discuss the case of $T_5^{-n}$, the parameters $\alpha_i$ are
those in expression \eqref{2-8orbits} which one obtains after setting
$  n_5=n, n_1= n_2=n_4=0$ and they satisfy
$\alpha_2=0, \alpha_3=0, \alpha_6=0$
as in equation \eqref{alpha632zero}.
As shown below \eqref{alpha632zero} we are left with
one Riccati equation for $q_1$ : $z q_{1\, z} = zq_1(z-q_2)+z  \alpha_1
+q_1 (1-\alpha_1-\alpha_5)$. Substituting  $\alpha_1=2-\mathsf{a}_4$ and $\alpha_5=-2n$.
we get
\begin{equation} z q_{1\, z} =zq_1(z-q_1)+z  \alpha_1
+q_1 (1-\alpha_1-\alpha_5)= zq_1(z-q_1)+z  (2-\mathsf{a}_4)
+q_1 (1-2+\mathsf{a}_4+2n) \, .
\label{T5q1}
\end{equation}
The solution is 
\begin{equation}
q_1(n,\mathsf{a}_4,z)= -\frac{n \mathsf{a}_4}{z} \frac{{\rm WhittakerW}
(-\frac12-\frac{\mathsf{a}_4}{4}
+\frac{n}{2}, \frac{n}{2}+\frac{\mathsf{a}_4}{4},- \frac{z^2}{2}))
}{{\rm WhittakerW}(\frac12-\frac{\mathsf{a}_4}{4}+\frac{n}{2}, 
\frac{n}{2}+\frac{\mathsf{a}_4}{4}, -\frac{z^2}{2})}
+\frac{z^2+2 n}{z}\, ,
\label{q1nalpha4}
\end{equation}
for which we find for $n=0,1,2$:
\begin{equation}
\begin{split}
q_1 (n=0,\mathsf{a}_4,z) &= z, \;\quad q_1 (n=1,\alpha_3,z)  =  z
\frac{-\mathsf{a}_4+z^2+2}{-\mathsf{a}_4+z^2}\\
p_1 (n=2,\mathsf{a}_4,z) &=
z \frac{-2\mathsf{a}_4+\mathsf{a}_4^2-2 z^2\mathsf{a}_4+z^4+4
z^2}{2\mathsf{a}_4+\mathsf{a}_4^2-2 z^2\mathsf{a}_4+z^4}\, ,
\label{q1T5}
\end{split}
\end{equation}
which are in agreement with results of acting with $T_5$ on 
solution given in \eqref{N6polynomialsols5}.
\end{exmp}

\section{Summary and Comments}
We identified rational solutions of the dressing chain
equations of even periodicity with points of an orbit generated 
by the fundamental shift operators acting on all first-order polynomial solutions.
It was described how additional B\"acklund 
transformation was needed to regularize those solutions that initially contained
a simple pole.

For those first-order polynomial solutions which contain neighboring $j_i$ and
 $j_{i+1}$ such that : $j_i+j_{i+1}=0$ for some $1 \le i \le N$
the action of some shift operators is not well-defined. 
Accordingly those shift operators  needed to be excluded in such cases 
and we have described the exclusion procedure in the paper. For orbits 
of the remaining well-defined shift operators
we showed how this structure for $N=4$ is responsible for a separate 
class of corresponding rational
solutions  (item III on page 
\pageref{page:items1}) of Painlev\'e V equation. 
We also showed how the rational solutions generated by a single shift
operator $T_i^n$  are expressed by Kummer/Whittaker polynomials with
arguments depending on integer $n$.

The advantage of the formalism we presented is that it is universal,
meaning that the derivation applies to all even-cyclic dressing chain systems
or equivalent  $A^{(1)}_{N-1}$ Painlev\'e equations as illustrated for the
case of $N=6$ in addition to the $N=4$ case. 

It is interesting to compare the derivation of elementary seed solutions for even-cyclic dressing chains 
with those encountered for odd-cyclic dressing chains. There are fundamental differences as the  $\alpha_i$
parameters are fixed and do not depend on arbitrary variables. Also in contrast to the even-cyclic dressing chains, 
the fundamental variables $j_i, i=1, ...,N$ of the odd-cyclic dressing chains that satisfy equations \eqref{dressingeqs}
  and 
the Painlev\'e variables $f_i, i=1,...,N$ are fully equivalent as the relation $f_i=j_i+j_{i+1}$  is reversible 
through expression $j_i =\frac12 \sum_{k=0}^{N-1} f_{i+k}$ for $N$ odd.
For example for $N=3$ one finds two elementary seeds solutions that can be written as $j_i=(z/2)(1,1,1)$, $\alpha_i=(1,1,1)$
and $j_i=(3z)/2) (-1,1,1)$,   $\alpha_i=3(0,1,0)$. It is well known that the rational solutions of the Painlev\'e IV equation can all be obtained by B\"acklund transformations from the above two seed solutions \cite{murata} whether expressee in terms of $J_i$ or $f_i$.

The natural next step, which we plan to pursue in the future,  
is to apply this framework to obtain closed determinant or 
special function expressions 
for rational solutions of all dressing chain
equations of even periodicity generated by combined shift
operators.

\vspace{5mm}

{\bf Acknowledgments}
This study was financed in part  by the Coordena\c{c}\~{a}o de Aperfei\c{c}amento de Pessoal de N\'ivel Superior - Brasil (CAPES) - Finance Code 001
(G.V.L.) and by CNPq and FAPESP (J.F.G. and A.H.Z.).

\appendix
\section{Derivation of $\pmb{A^{(1)}_5}$ Painlev\'e equations for
$\pmb{N=6}$ dressing chain}

The dressing chain equations \eqref{dressingeqseven} can be rewritten
entirely in terms of $f_i$ without any references 
to $j_i$ after inserting the value for $\Psi$. It needs to be
emphasized that such elimination of $j_i$ variables while expressing
the dressing chain equations in terms of $f_i$ requires inserting the
definition of $\Psi$ from \eqref{psidef} into equations \eqref{dressingeqseven}. 
Such substitution of $j_i$ by $f_i$ would not work with 
equations \eqref{dressingeqs} for $N=$ even.

For $N=6$ such procedure yields $A^{(1)}_5$ Painlev\'e equations:
\begin{equation}
\begin{split}
z f_{1\, z} &= f_1 f_3(f_2-f_4-f_6)+f_1 f_5 (f_4+f_2-f_6) + z \alpha_1+
f_1(1-\alpha_1-\alpha_3-\alpha_5) \, ,\\
z f_{2\, z} &= f_2 f_4(f_3-f_1-f_5)+f_2 f_6 (f_3-f_1+f_5) + z \alpha_2-
f_2 (1-\alpha_1-\alpha_3-\alpha_5)\, , \\
z f_{3\, z} &= f_1 f_3(f_4-f_2+f_6)+f_3 f_5 (f_4-f_2-f_6) + z \alpha_3+
f_3(1-\alpha_1-\alpha_3-\alpha_5)\, , \\
z f_{4\, z} &= f_2 f_4(f_1-f_3+f_5)+f_4 f_6 (f_5-f_1-f_3) + z \alpha_4-
f_4 (1-\alpha_1-\alpha_3-\alpha_5) \, ,\\
z f_{5\, z} &= f_1 f_5(f_6-f_2-f_4)+f_3 f_5 (f_6+f_2-f_4) + z \alpha_5+
f_5(1-\alpha_1-\alpha_3-\alpha_5) \, ,\\
z f_{6\, z} &= f_2 f_6(f_1-f_3-f_5)+f_4 f_6 (f_1+f_3-f_5) + z \alpha_6
-f_6(1-\alpha_1-\alpha_3-\alpha_5)\, , \\
\end{split}
\label{N6NY}
\end{equation}
with $z=f_1+f_3+f_5=f_2+f_4+f_6$.

\end{document}